\documentclass[twocolumn,twocolappendix,iop,numberedappendix,appendixfloats]{openjournal}

\usepackage[utf8]{inputenc}

\usepackage[british]{babel}

\usepackage{amsmath}
\usepackage{amssymb}
\usepackage{amsfonts}
\usepackage{graphicx}
\usepackage{bm}
\usepackage{natbib}
\usepackage[colorlinks,allcolors=blue]{hyperref}

\usepackage{subfigure}
\usepackage{xspace}
\usepackage{multirow}
\usepackage{caption}
\usepackage{subcaption}
\usepackage{fontawesome}
\usepackage{placeins}

\newcommand{\IAEmu}{\texttt{IAEmu}\xspace}
\newcommand{\halotoolsia}{\texttt{halotools-IA}\xspace}
\newcommand{\githubmaster}{\href{https://github.com/snehjp2/IAEmu}{\faGithub}\xspace}

\begin{document}

\journalinfo{The Open Journal of Astrophysics}

\shorttitle{Learning Galaxy IA Correlations}
\shortauthors{S. Pandya et al.}

\title{IAEmu: Learning Galaxy Intrinsic Alignment Correlations}

\author{Sneh Pandya$^{1,2}$}
\author{Yuanyuan Yang$^{1,3}$}
\author{Nicholas Van Alfen$^{1}$}
\author{Jonathan Blazek$^{1}$}
\author{Robin Walters$^{3}$}

\affiliation{$^{1}$Department of Physics, Northeastern University, Boston, MA 02115, USA}
\affiliation{$^{2}$NSF AI Institute for Artificial Intelligence and Fundamental Interactions (IAIFI)}
\affiliation{$^{3}$Khoury College of Computer Sciences, Northeastern University, Boston, MA 02115, USA}
\thanks{$^\star$ E-mail: \nolinkurl{pandya.sne@northeastern.edu}}

\begin{abstract}
The intrinsic alignments (IA) of galaxies, a significant contaminant in weak lensing analyses, arise from correlations in galaxy shapes driven by gravitational tidal interactions and galaxy formation processes. Understanding IA is therefore essential for deriving accurate cosmological inferences from weak lensing surveys. However, most IA modeling relies on a combination of perturbative approaches, which cannot describe nonlinear scales, and expensive simulation-based approaches. In this work, we introduce \IAEmu, a neural network-based emulator designed to predict the galaxy position-position ($\xi$), position-orientation ($\omega$), and orientation-orientation ($\eta$) correlation functions, and their associated uncertainties, using halo occupation distribution (HOD)-based mock galaxy catalogs. Compared to the simulated catalogs, \IAEmu exhibits an approximately $3\%$ average error for $\xi$ and $5\%$ for $\omega$, while capturing the stochasticity in $\eta$, avoiding overfitting this inherently noisier statistic. Importantly, the emulator also provides aleatoric and epistemic uncertainties, which when analyzed jointly, can help identify regions in parameter space where \IAEmu's predictions may be less reliable. Furthermore, we demonstrate the model’s generalization to a non-HOD based signal by fitting alignment parameters from the \textsc{tng300} hydrodynamical simulations. Since \IAEmu is a fully differentiable neural network, it enables approximately a $10{,}000\times$ speed-up in mapping HOD parameters to correlation functions when deployed on a GPU, compared to conventional CPU resources. This substantial acceleration also facilitates solving inverse problems more efficiently by supporting gradient-based sampling algorithms. As such, \IAEmu offers an efficient and promising surrogate model for halo-based galaxy bias and IA modeling with the potential to expedite model validation in Stage IV weak lensing surveys. \githubmaster
\end{abstract}

\keywords{
Weak Gravitational Lensing, Intrinsic Alignment, Machine Learning
}

\maketitle

\section{Introduction}

Weak lensing, a subtle yet rich effect that maps the distribution of dark matter and measures cosmic structure growth, is a key cosmological probe for the Legacy Survey of Space and Time (LSST, \cite{lsst}) of the Vera C. Rubin Observatory, as well as for the \emph{Roman} \citep{roman} and \emph{Euclid} \citep{euclid} missions.
These next-generation surveys will be incredibly powerful, achieving sub-percent levels of statistical precision.
The intrinsic alignment (IA) of galaxies, a result of their interactions with large-scale structure and other galaxies, can contaminate weak lensing measurements, necessitating accurate and efficient IA modeling in preparation for future analyses \citep[e.g.][]{TROXEL20151,krause16,Blazek_2019,fortuna21a,Hoffmann_2022,secco22,campos23,samuroff24, Paopiamsap_2024}.

Intrinsic alignment (IA) modeling has traditionally relied on analytic approaches, such as perturbation theory \citep[e.g.][]{Lamman_2024,Bridle_2007,Blazek_2015,Blazek_2019,vlah_2020,vlah_2021,maion_2023,bakx_2023,chen_2023}.
However, these analytic models often struggle to accurately capture nonlinear effects. 
In cosmology, simulation-based approaches that account for both gravitational and baryonic effects have provided profound insights into cosmological evolution \citep[e.g.][]{camels,nelson2021illustristng,pillepich18,delgado_2023}.
These methods have the potential to better capture IA effects compared to purely analytic models. 
Magnetohydrodynamic simulations, often referred to as ``hydro'' simulations, incorporate baryonic effects but exhibit significant variance in their predictions depending on the simulation suite and the sub-grid physics models employed at sub-parsec scales.
This variance includes disagreement on IA effects measured in different simulation suites \citep[e.g.][]{tenneti16,samuroff21}.
However, the volumes required for modern cosmological analyses are prohibitively large for hydrodynamical simulations.
Given these challenges, gravity-only $N$-body simulations, such as \emph{AbacusSummit} \citep{abacussummit} and \emph{Quijote} \citep{quijote}, provide a cost-effective and general alternative.
These simulations avoid the need to specify sub-parsec-scale baryonic physics, but inherently lack galaxy formation and evolution processes.
To bridge this gap, various halo occupation distribution (HOD) models have been employed to populate halos from $N$-body simulations with galaxies.
This approach has been extended to include galaxy populations that exhibit correlated alignments \citep[e.g.][]{Joachimi_2013,Hoffmann_2022,vanalfen_2023}. 
At the scales required for cosmological inference, emulators built on these $N$-body and HOD frameworks are essential for rapidly generating accurate mock galaxy catalogs across large parameter spaces.

Despite its utility, HOD modeling can still be computationally demanding, as it requires the generation of extensive galaxy catalogs from halo catalogs and the application of estimators to extract IA and clustering signals. To reduce computational cost and time associated with IA modeling, surrogate modeling and emulation based on existing simulations represents a promising avenue of research. One approach that can offer both precision and efficiency is deep learning (DL), by training neural network (NN) surrogates to accelerate numerical simulations.
NNs have seen many different scientific applications, including in cosmology (see \citet{dvorkin2022machine} for a review), with the availability of large datasets and powerful GPU-driven computation. 
They not only provide new insights, but also have the potential to accelerate numerous analyses when deployed on GPUs. The benefits of NN-based surrogate models are not exclusive to forward modeling, as the differentiability of such models can also be exploited in accelerating inverse problems using differentiable sampling techniques.

In this work, we introduce an NN-based approach to emulating HOD simulations that models both galaxy bias and IA statistics, referred to as \IAEmu. 
\IAEmu is the first NN-based model to directly predict galaxy position and shape correlations from HOD parameters, eliminating the need to rerun simulations or generate explicit galaxy catalogs.
This approach offers a significant speed advantage over traditional HOD-based modeling.
Additionally, \IAEmu successfully models galaxy shape statistics, whose stochasticity is dominated by galaxy shape noise, as discussed in \citet{vanalfen_2023}. \IAEmu successfully captures the mean behavior of these noisier statistics, which would otherwise require multiple realizations of the underlying HOD. It also estimates galaxy shape noise (aleatoric uncertainty) and quantifies its own epistemic uncertainty -- reflecting uncertainty in the predicted correlation amplitudes -- primarily due to limited training data. \IAEmu's uncertainty estimates enable one to assess the reliability of these predictions and further enable error propagation in modeling pipelines that incorporate \IAEmu. We further show the benefits of accelerated parameter inference (i.e., inverse problems) using gradient-based sampling techniques with \IAEmu, exploiting the fact that NNs are differentiable models.

\textbf{Related Work.} Several previous works have constructed simulation-based emulators for cosmological statistics, with a focus on matter or galaxy density.
\cite{Zhai_2019} constructed Gaussian process-based emulators based on the \textsc{aemulus} Project's $N$-body simulations for nonlinear galaxy clustering. 
\cite{Kwan_2023} similarly used a Gaussian process-based emulator, HOD modeling, and the \textsc{Mira-Titan} Suite of $N$-body simulations to predict galaxy correlation functions, building on earlier work from the same group \citep{2010ApJ...713.1322L}.
The \textsc{Bacco} simulation project (\citealt{2021arXiv210414568A}, \citealt{2021MNRAS.506.4070A}) built NN emulators to include nonlinear and baryonic effects from simulations.
These projects emulate various cosmological statistics from simulations, but do not include IA. \cite{Jagvaral_2022}, \cite{ jagvaral24_arxiv}, and \cite{2023arXiv231211707J} developed generative models trained on the \textsc{tng100} simulation \citep{nelson2021illustristng} to emulate IA in hydrodynamic simulations, but these models do not emulate statistics. 
Our work is the first to emulate galaxy-IA correlation statistics using simulated galaxy catalogs.

\textbf{Paper Organization.} This paper is organized as follows. Section 2 provides a background on the HOD simulation and correlation function estimators, as well as the procedure for generating and cleaning the training and test data.
Section 3 introduces the \IAEmu architecture and the process for training \IAEmu.
In Section 3, we also analyze the generalization performance of \IAEmu on held-out data and characterize the quality of predictions based on the predicted aleatoric and epistemic uncertainty.
Finally, in Section 4, we validate the out-of-distribution (OOD) performance of \IAEmu on a non-HOD-based signal from the \textsc{tng300} suite of simulations \citep{Nelson_2015, Pillepich_2017, Springel_2017, Nelson_2017, Naiman_2018, Marinacci_2018} by obtaining a posterior on alignment parameters, which is compared with the HOD-based approach. We summarize our main results in Section 5.

\begin{figure*}
    \centering
    \includegraphics[width=\textwidth]{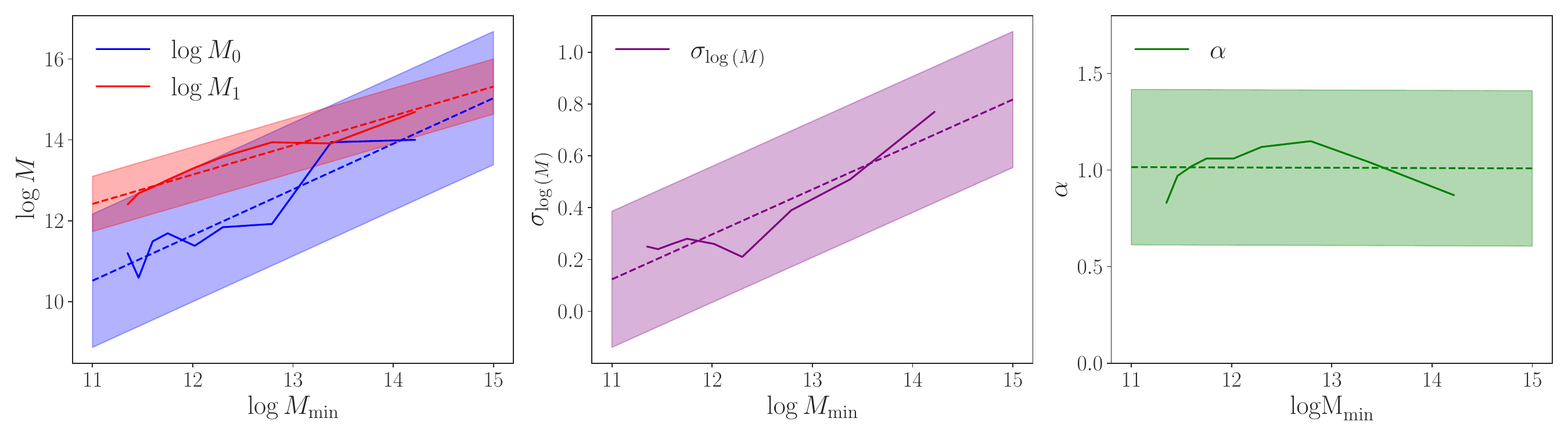}
    \caption{Ranges of HOD parameters used in generating the training data from \halotoolsia. We generate uniform random values for the four occupation parameters, excluding $\log M_{\text{min}}$. These values are based on a linear relationship with $\log M_{\text{min}}$, serving as a central line. The range for random values extends $4 \cdot \text{RMSE}$ surrounding this line. To clarify the visualization, $\sigma_{\log (M)}$ is displayed separately from other mass variables. Each panel presents published data from \citet{Zheng_2007} as a solid line, while the dotted line of the same color illustrates the linear fit to $\log M_{\text{min}}$, with the shaded area indicating the range for uniform random value selection for each parameter. Not shown here are the two alignment parameters, $\mu_{\text{cen}}$ and $\mu_{\text{sat}}$, which both vary uniformly on the range $[-1, 1]$ with no relation to these five occupation parameters.}
    \label{fig:params}
\end{figure*}

\section{Dataset: Halo Occupation Distribution}
\label{sec:HOD}

This section outlines the basics of the halo occupation distribution, the estimators used by {\tt halotools} to measure correlations, and the generation of the data on which \IAEmu was trained.

\subsection{HOD Background \& Estimators}
\label{sec:HOD_estimators}

Given a catalog of dark matter halos, we generate a galaxy catalog using an HOD model.
This model consists of several interconnected components: (1) an occupation component, which populates halos with galaxies, (2) a phase space component, which determines the spatial distribution of galaxies within halos, and (3) an alignment component, which models galaxy intrinsic alignments.
The \texttt{halotools} \cite{Hearin_2017} package constructs these HOD-based galaxy catalogs following this framework.
Specifically, it employs the halo model \citep{cooray_2002, asgari_2023} along with alignment models introduced in \citet{vanalfen_2023}, providing a flexible approach for generating mock galaxy catalogs while simultaneously tracking intrinsic alignments.
We refer to this extension of \texttt{halotools}, which incorporates IA information, as \halotoolsia \citep{VanAlfen2025}.
This structure enables the rapid generation of multiple galaxy catalogs using consistent occupation, phase space, and alignment parameters. Depending on the chosen HOD parameters, a given halo may or may not host a central galaxy -- the most massive galaxy residing at the halo's center. 
Additionally, halos may contain satellite galaxies, which are distributed throughout the halo.

\textbf{Correlation Estimators.} To measure the correlations in these catalogs, \halotoolsia uses the estimators in Equations \eqref{eq:xi-estimator}-\eqref{eq:eta-estimator} for the position-position ($\xi$), position-orientation ($\omega$), and orientation-orientation ($\eta$) correlations, respectively. 
The $\xi$ correlation is defined as
\begin{equation}
    \label{eq:xi-estimator}
    \xi(r) = \left\langle\frac{n(r)}{\bar{n}(r)}\right\rangle - 1,
\end{equation}
where $n(r)$ is the number of galaxies separated by distance $r$, and $\bar{n}(r)$ is the expected number of galaxies separated by distance $r$ for a random distribution.
This equation is simpler than the Landy-Szalay estimator \citep{landy_1993} and may be suboptimal in some cases \citep{sukhdeep_2017}, since it omits the random-pair corrections that reduce variance and account for survey geometry that are present in Landy-Szalay, making it more sensitive to noise and boundary effects.
However, due to the periodic nature of the simulation box, \halotoolsia can use analytical randoms, mitigating much of this suboptimality.
This estimator is also computationally faster and is sufficient for our HOD models. The $\omega$ correlation is defined as
\begin{equation}
\label{eq:omega-estimator}
\omega(r) = \langle |\hat{e}({\bf x}) \cdot \hat{r}|^2 \rangle -\frac{1}{3} \;, 
\end{equation}
and quantifies how the orientation of a galaxy at a position $\bf{x}$ is aligned with the positions of other galaxies at a distance $\bf{r}$.
If $\omega$ is positive, the orientation tends to align with the direction to nearby galaxies; if negative, it tends to be perpendicular.
Similarly, the $\eta$ correlation is defined as
\begin{equation}
\label{eq:eta-estimator}
\eta(r) = \langle |\hat{e}({\bf x}) \cdot \hat{e}({\bf x}+{\bf r})|^2 \rangle -\frac{1}{3},
\end{equation}
and measures how similarly two galaxies at positions $\bf{x}$ and 
$\bf{x} + \bf{r}$ are oriented.
A positive $\eta$ indicates that the orientations tend to be aligned, while a negative value means they tend to be perpendicular.
For both $\omega$ and $\eta$, ${\bf x}$ is the position vector of a given galaxy, ${\bf r}$ is the separation vector between two galaxies, $\hat{r}$ is the unit vector of the separation vector ${\bf r}$, and $\hat{e}$ is the galaxy orientation unit vector that specifies the intrinsic orientation of each galaxy's major axis.
The factor of $1/3$ in these equations accounts for the fact that
\begin{equation}
    \frac{1}{4\pi} \int_0^{2\pi} \int_0^\pi \cos^2 \theta \sin \theta \, \mathrm{d}\theta \, \mathrm{d}\phi = \frac{1}{3} \; ,
\end{equation}
where integrating $\cos^2{\theta}$ over a sphere corresponds to the case of random alignments.

In this work, correlation functions are measured for simulated galaxies across 20 $r$ bins, evenly spaced in logarithmic scale, between a minimum separation of \( 0.1 \, h^{-1} {\rm Mpc} \) and a maximum separation of \( 16 \, h^{-1} {\rm Mpc} \).
In future work, the maximum range of this correlation could be extended.
However, for this dataset, we chose this maximum separation because the number of galaxies $n$ increases with $r$, and the computational cost of measuring correlations scales as \( \mathcal{O}(n) \log (n) \).
In general, galaxies at $r \leq 1 \, h^{-1} {\rm Mpc}$ are considered to be in the ``1-halo regime'' (galaxies within the same halo) and galaxies outside this range are in the ``2-halo regime'' (galaxies residing in separate halos). 

\subsection{Dataset Generation}
\label{sec:arch}

\begin{table*}[t]
\centering

\caption{\textsc{\textsc{Bolshoi-Planck}} simulation parameters.}
\begin{tabular}{lccccccc}
\hline
Simulation & Particle Mass & $\Omega_{m,0}$ & $\sigma_8$ & $n_s$ & $h$ & $L_{\rm box}$ & $z$ \\
 & ($h^{-1}M_\odot$) &  &  &  &  & ($h^{-1}$Mpc) &  \\
\hline
\textsc{\textsc{Bolshoi-Planck}} & $\sim 10^8$ & 0.30711 & 0.82 & 0.96 & 0.70 & 250 & 0 \\
\hline
\end{tabular}
\label{tab:bolshoi}
\end{table*}

To train \IAEmu, we generate galaxy catalogs using \halotoolsia, incorporating seven HOD and IA parameters derived from an existing dark matter halo catalog that is consistent with a realistic cosmology.
We use dark matter catalogs from the \textsc{Bolshoi-Planck} simulations, which are available directly through \halotoolsia for this purpose \citep{Klypin_2011}.
The \textsc{Bolshoi-Planck} simulation is a $250 \; h^{-1} \text{Mpc}$ box, with correlations in this work measured up to $16 \; h^{-1} \text{Mpc}$.
Simulation parameters for \textsc{Bolshoi-Planck} can be found in Table \ref{tab:bolshoi}.
We populate halos with galaxies following occupation equations from \citet{Zheng_2007}.
To choose physically plausible values for the five occupation parameters used by these two models, we select the best-fit HOD parameter values for the Sloan Digital Sky Survey (SDSS) sample from Table 1 of \citet{Zheng_2007}.
Further details of how we employ these occupation methods as well as a discussion of the phase space and alignment models are given in Appendix \ref{sec:appendix-hod-setup}.

The five occupation parameters are: $\log{M_{\text{min}}}$, $\log{M_{0}}$, $\log{M_{1}}$, $\alpha$, and $\sigma_{\log{M}}$.
The parameters $\log{M_{\text{min}}}$, $\log{M_{0}}$, and $\log{M_{1}}$ control the relationship between dark matter halo masses and the likelihood of hosting central and satellite galaxies in the HOD model.
Specifically, $\log{M_{\text{min}}}$ defines the minimum halo mass required to host a central galaxy, $\log{M_{0}}$ sets the mass scale associated with the suppression of the satellite galaxy occupation, and $\log{M_{1}}$ determines the amplitude of the satellite occupation profile.
The number of galaxies in a given catalog ranges from $10^5$ to $10^6$, with the average number decreasing with larger $\log M_{\text{min}}$.
The parameter $\alpha$ describes the asymptotic slope of satellite occupation at high halo masses, while $\sigma_{\log{M}}$ characterizes the width of the transition between halos that do and do not host central galaxies.
Figure \ref{fig:params} shows the regions from which four of the five occupation model parameters are drawn.

The two alignment parameters, $\mu_{\rm cen}$ and $\mu_{\rm sat}$, govern the shape of the Dimroth-Watson distribution from which galaxy misalignments are sampled, as introduced in \citet{vanalfen_2023}.
More specifically, an alignment parameter value of 0 corresponds to a uniform distribution in $\cos{(\theta)}$, where $\theta$ is the galaxy misalignment angle, indicating randomly oriented galaxies.
Values approaching $1$ indicate perfectly aligned galaxies, while values approaching $-1$ correspond to perpendicular alignments.

To generate training data, we generate evenly spaced values of $\log{M_{min}}$ within the range $[11,15]$, covering the typical halo mass scales that host galaxies.
For each of these points, we draw a value for each of the other four occupation parameters uniformly from a region $\pm 4 \cdot {\rm RMSE}$ around the linear fit to $\log{M_{\text{min}}}$, where RMSE refers to the root mean squared error between the fiducial SDSS values of each occupation parameter from \citet{Zheng_2007} and their corresponding values predicted by the linear fit. 
The two alignment parameters, $\mu_{\rm cen}$ and $\mu_{\rm sat}$, are each sampled uniformly on $[-1,1]$. 
Earlier iterations of the dataset employed Latin hypercube sampling to fully sample the parameter space; however, this frequently resulted in galaxy samples that were unrealistic. 
For this reason, we restricted our sampling to values lying near the empirical trends of the SDSS-based HOD fits, which ensures that the resulting correlations remain physically plausible while still spanning a representative range of parameter space.

Despite the constrained sampling strategy, it was observed that certain input configurations could lead to an absence of galaxy pairs in specific separation bins, resulting in \texttt{NaN} values in the correlation functions. 
This issue frequently arises at small scales, or when the values of $\log{M_{\text{min}}}$ are sufficiently large, making the halos that host galaxies rare. 
To address this, corresponding galaxy catalogs were removed. 
Additionally, as a further screening measure, we impose a restriction on input configurations that yield $\xi/\xi_{\text{DM}}$ values exceeding 100, as these are deemed unphysical. 

With these input parameter values, we generate galaxy catalogs using \halotoolsia and measure the three correlations described in Section \ref{sec:HOD_estimators}.
As HOD modeling is inherently stochastic, we generate 10 realizations of a galaxy catalog for each given set of input values for training.
The multiple realizations can enable \IAEmu to distinguish the signal from the shape noise of the data, and they later serve to quantify the performance of \IAEmu for the noisier correlations.
We note that the sample variance does also contribute to the variance in the correlations, but it is always subdominant to the shape noise (see Appendix D of \citet{vanalfen_2023} and Equation A3 of \citet{Chisari_2013}).
Thus, we can capture most of the statistical variance by re-aligning galaxies through these extra realizations.
The final dataset has $110,526$ parameter choices, with 10 realizations, for a total of $1,105,260$ entries.
These are split into a 70\% train, 10\% validation, and 20\% test set with unique input parameters in each subset.
The training data was generated using a combination of 2.4 GHz Intel E5-2680 CPUs and 2.1 GHz Intel Xeon Platinum 8176 CPUs.
The simulations were parallelized across 150 cores, split evenly to allow simultaneous calculation of the correlation functions. 

\section{The IAEmu Model}

\begin{figure*}
    \centering
    \includegraphics[width=\textwidth]{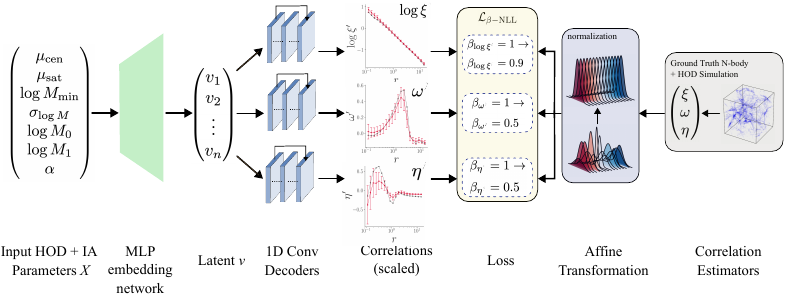}
    \caption{Model Pipeline. The HOD input model parameters are normalized before entering the 7-layer deep multilayer perceptron (MLP) embedding network. The embedding network expands the dimensionality of the input before a bottleneck latent space that transitions to the decoder stage, which features seven 1D convolutional layers which learn the individual local correlations present in the output correlation functions, $\log \xi$, $\omega$, and $\eta$. Both the embedding network and decoder feature residual connections to aid the convergence of \IAEmu during training. \IAEmu is trained using the $\beta$-NLL loss \citep{seitzer2022pitfallsheteroscedasticuncertaintyestimation} with a 100 epoch warm-up period corresponding to mean-squared-error optimization before re-introducing aleatoric uncertainties into the optimization. The generated correlation functions are then re-scaled back to their original values. A detailed description of the model training procedure is shown in Appendix \ref{sec:appendix-training}. $N$-body simulation visualization in the right panel is from \citep{nbody}.}
    \label{fig:pipeline}
\end{figure*}

In this section, we summarize the \IAEmu architecture and training procedure.

\subsection{Model Architecture}

Our objective is to construct a neural network (NN) that replicates the mapping between HOD simulations and correlation function estimators.
Specifically, the NN will take a 7-dimensional input vector of galaxy HOD and IA parameters, as described in Section \ref{sec:HOD} and illustrated in Figure \ref{fig:pipeline}, and predict the correlation functions \(\xi(r)\), \(\omega(r)\), and \(\eta(r)\) across 20 $r$ bins.
We represent each correlation function by a vector recording the value for 20 evenly spaced values of $r$.
Additionally, the model directly outputs predictions of the aleatoric uncertainties $\sigma^{\text{aleo}}$ on the correlation amplitudes.
This allows us to capture the stochastic nature of HOD modeling through a mean-variance estimation (MVE) training procedure \citep{MVE}.
Separately, we use Monte Carlo dropout to track the epistemic uncertainties $\sigma^{\text{epi}}$ inherent in the NN model.
These can arise from limited training data or architecture misspecification.
Both types of uncertainties are useful for analyzing $\omega(r)$ and $\eta(r)$ performance, which are inherently noisier statistics due to the significant effects of galaxy shape noise in correlations \citep{Bernstein_2002}.
Mathematically, the task mapping is a function 
\begin{align*}
f_{\phi} \colon \mathbb{R}^7 &\rightarrow \mathbb{R}^{2 \times 20} \times \mathbb{R}^{2 \times 20} \times \mathbb{R}^{2 \times 20}
\end{align*}
where $f_\phi$ maps the input $X$ to a set of mean and aleatoric uncertainty pairs:
\begin{align*}
X \mapsto (
\underbrace{[\mu_\xi, \sigma_\xi^{\text{aleo}}]}_{\in \mathbb{R}^{2 \times 20}}, \quad
\underbrace{[\mu_\omega, \sigma_\omega^{\text{aleo}}]}_{\in \mathbb{R}^{2 \times 20}}, \quad
\underbrace{[\mu_\eta, \sigma_\eta^{\text{aleo}}]}_{\in \mathbb{R}^{2 \times 20}}
).
\end{align*}
We implement $f_\phi$ as an NN called \IAEmu using \texttt{PyTorch} \citep{paszke2019pytorchimperativestylehighperformance}.
The \IAEmu architecture includes a fully connected embedding network and three 1D convolutional NN decoder heads, trained using a multitask learning approach as shown in Figure \ref{fig:pipeline}.

The embedding network contains five fully connected linear layers, each followed by batch normalization and \texttt{LeakyReLU} activation \citep{xu2015empirical}. 
Residual connections link the second and third layers using a linear projection to match dimensions, and the third and fourth layers by directly adding the layer outputs, which improves information flow and gradient stability \citep{he2015deepresiduallearningimage}.
The embedding network increases the size of the input vector $X \in \mathbb{R}^7$ layer-by-layer to a $256$-dimensional latent feature, which is then mapped through a final bottleneck layer to a 128-dimensional latent vector $v$.
To mitigate overfitting, we incorporate dropout \citep{JMLR:v15:srivastava14a} into the \IAEmu architecture during training. Additionally, as detailed later, we leverage dropout to estimate the epistemic uncertainty associated with the model's parameters using the Monte Carlo dropout technique \citep{gal2016dropout}.

The decoders each contain seven 1D convolutional decoder layers.
Each decoder first takes the output of the embedding network, a feature vector $v$ of size 128, and maps it into an expanded feature space.
This expanded feature vector is then reshaped to create a multichannel 1D feature map, enabling the decoder to utilize 1D convolution to spatially transform the latent representation.
Each layer has batch normalization, \texttt{LeakyReLU} activation, and dropout.
Residual connections are introduced by adding the output of the second convolutional layer to the output of the third layer and by adding the output of the fifth layer to the output of the sixth layer.
Each decoder gradually downsamples the latent representation $v$ and finally outputs a 2-channel 1D signal as a tensor of shape $2\times20$, where the 2 channels represent the correlation amplitudes and variances of the correlation function, respectively.
To ensure variances are strictly positive, they are passed through a \texttt{softplus} activation in the output layer.

The \IAEmu design serves a dual purpose: it facilitates vector-to-sequence conversion through the convolution of encoded representations from the embedding network and, within our multitask framework, enables separate forward paths to isolate features unique to each individual correlation estimator. 

\subsection{Training}
\label{subsec:training}

We now describe the training procedure for \IAEmu. 
We normalize each feature within the 7-dimensional input vector $X \in \mathbb{R}^7$ such that the overall distribution of each component of $X$ has a mean of 0 and unit variance. That is, each individual feature $x$ (i.e., a single component of $X$) undergoes the transformation: 
\begin{equation}
\label{eqn:scaler}
z = \frac{x - \mu_x}{\sigma_x} \; ,
\end{equation}
where $\mu_x$ and $\sigma_x$ are the mean and standard deviation of the respective feature across the entire training dataset.
This is known as $z$-score standardization; it is an \emph{affine} transformation and is thus easily invertible.

We are interested in predicting three sequences, each of length 20, corresponding to the correlation functions $\xi(r)$, $\omega(r)$, and $\eta(r)$ for $0.1 \;h^{-1} \text{Mpc} < r < 16 \;h^{-1} \text{Mpc}$.
Since these correlations exhibit different magnitudes and characteristics, each correlation function is also standardized separately for the training of \IAEmu.
This ensures that each correlation is scaled to have a mean of 0 and unit variance across all bins.
Without this, the loss landscape would be unevenly influenced by the differing magnitudes of the correlation functions.
For example, $\xi(r)$ can exhibit strong correlations at low values of $r$, reaching amplitudes on the order of $10^4$ or higher.
In contrast, $\omega(r)$ and $\eta(r)$ exhibit amplitudes several orders of magnitude smaller than $\xi(r)$, and can also frequently take on negative values.
Applying separate standardization to each correlation function ensures that all three contribute equally to the loss landscape during training.
Since $\xi$ can vary over several orders of magnitude, we take its logarithm before $z$-score standardization.
This transformation reduces skewness and can help mitigate the dominance of high-magnitude correlations in the standardization process.
We thus denote the \IAEmu predicted correlations as $\log \hat{\xi}(r)$, $\hat{\omega}(r)$, and $\hat{\eta}(r)$. This standardization additionally applies to the \IAEmu predicted aleatoric uncertainties: $\widehat{\sigma_{\log \xi}^{\text{aleo}}}$, $\widehat{\sigma_{\omega}^{\text{aleo}}}$, and $\widehat{\sigma_{\eta}^{\text{aleo}}}$, as well as to the epistemic uncertainties: $\widehat{\sigma_{\log \xi}^{\text{epi}}}$, $\widehat{\sigma_{\omega}^{\text{epi}}}$, and $\widehat{\sigma_{\eta}^{\text{epi}}}$.
All presented results are for rescaled correlations and uncertainties, with the rescaling transformations given in Appendix \ref{sec:appendix-rescaling}.

To predict the mean and variance of the values of the correlation function, we use the $\beta$-NLL loss from \citep{seitzer2022pitfallsheteroscedasticuncertaintyestimation}, which is defined as
\begin{align}
\mathcal{L}_{\beta\text{-NLL}} 
= \mathbb{E}_{X,Y} \bigg[
    & \hat{\sigma}^{2\beta}(X) \Big(
    \frac{1}{2} \log \hat{\sigma}^2(X) \notag \\
    & + \frac{(Y - \hat{\mu}(X))^2}{2\hat{\sigma}^2(X)} + C \Big)
\bigg]
\end{align}

This is similar to Gaussian-NLL loss \citep{MVE}, defined
\begin{equation}
\label{eq:NLL}
\mathcal{L}_{\text{NLL}} = \mathbb{E}_{X,Y} \left[ \frac{1}{2} \log \hat{\sigma}^2(X) + \frac{(Y - \hat{\mu}(X))^2}{2 \hat{\sigma}^2(X)} + C \right] \;,
\end{equation}
where $X$ denotes the input data vector, $\hat{\mu}(X)$ and $\hat{\sigma}^2(X)$ the model predictions at an individual bin, $Y$ the ground truth label, and $C$ a normalization constant.
The numerator of the second term in Equation \ref{eq:NLL} is the typical mean-squared-error loss, used when the model only outputs a point estimate approximating the mean of the distribution.
One drawback of the Gaussian NLL loss is that the model can become stuck in local minima in the loss landscape during training.
This results in a prediction with an incorrect mean and high variance.
However, by adjusting $\beta$ appropriately, this risk can be reduced. The utility of the $\beta$-NLL loss can be seen in the gradients:
\begin{align}
\nabla_{\hat{\mu}} \mathcal{L}_{\beta\text{-NLL}}(\theta) 
  &= \mathbb{E}_{X,Y} \left[ \frac{\hat{\mu}(X) - Y}{\hat{\sigma}^{(2 - 2 \beta)}(X)} \right] \\
\nabla_{\hat{\sigma}^2} \mathcal{L}_{\beta\text{-NLL}}(\theta) 
  &= \mathbb{E}_{X,Y} \left[ \frac{\hat{\sigma}^2(X) - (Y - \hat{\mu}(X))^2}{2 \hat{\sigma}^{(4 - 2 \beta)}(X)} \right] \;.
\end{align}

The $\beta$ parameter allows one to interpolate between Gaussian-NLL in the limit that $\beta \rightarrow 0$, and standard MSE in the limit that $\beta \rightarrow 1$.
This loss has the benefit of allowing one to encode the contribution of the mean prediction to the loss, to discourage local minima with poor mean predictions and large variances.
It was empirically found in \citet{seitzer2022pitfallsheteroscedasticuncertaintyestimation} that a value of $\beta = 0.5$ generally performs best.
However, we explore different values of $\beta$ and introduce a warm-up period of $\ell'$ epochs to enable individualized training for each correlation.
The total loss function during training at epoch $\ell$ is:
\begin{align}
\mathcal{L}(\ell; \theta) = 
\begin{aligned}[t]
    & 
    \begin{cases}
        \begin{aligned}
            & \mathcal{L}_{\beta\text{-NLL}}^\xi(\theta, 1.0)
            + \mathcal{L}_{\beta\text{-NLL}}^\omega(\theta, 1.0) \\
            & + \mathcal{L}_{\beta\text{-NLL}}^\eta(\theta, 1.0)
        \end{aligned}
        & \text{if } \ell < \ell' \\[2mm]
        \begin{aligned}
            & \mathcal{L}_{\beta\text{-NLL}}^\xi(\theta, 0.9)
            + \mathcal{L}_{\beta\text{-NLL}}^\omega(\theta, 0.5) \\
            & + \mathcal{L}_{\beta\text{-NLL}}^\eta(\theta, 0.5)
        \end{aligned}
        & \text{if } \ell \geq \ell'
    \end{cases}
\end{aligned}
\end{align}
where we set $\beta_\xi = 0.9$ after the warm-up as this is a higher-signal correlation.
Further details regarding the training hyperparameters can be found in Appendix \ref{sec:appendix-training}.

\section{Results}

\begin{figure*}
    \centering
    \includegraphics[width=\textwidth]{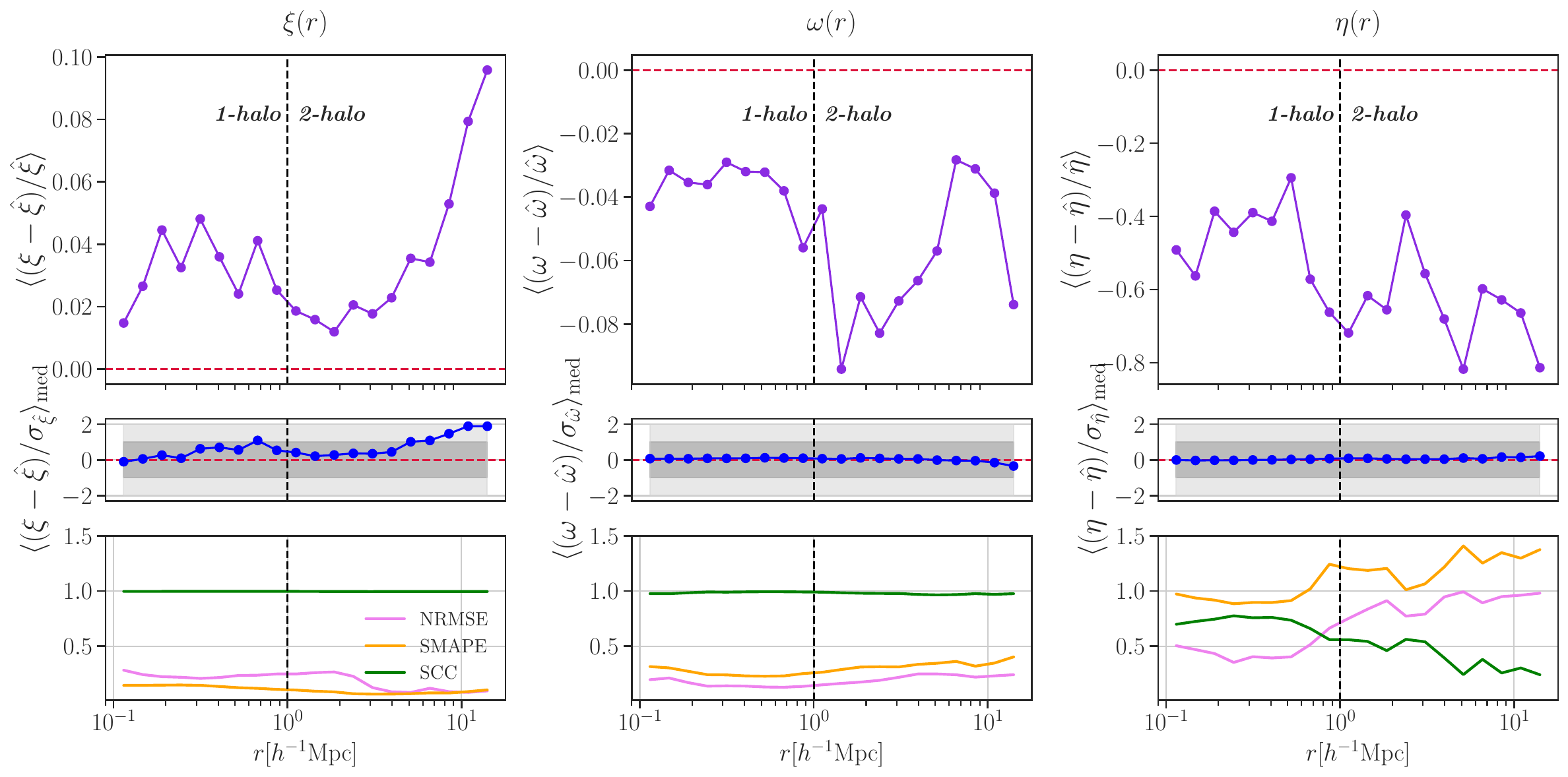}
    \caption{\textbf{Top:} Average fractional error 
    for the position-position ($\xi$), position-orientation ($\omega$), and orientation-orientation ($\eta$) correlation function predictions in the test set shown in purple. \textbf{Middle:} Median residuals of the test set predictions, expressed in units of the standard deviation of the ground truth data, $\hat{\sigma}$, obtained from 10 realizations used to construct the dataset shown in blue. \textbf{Bottom:} Per-bin Spearman correlation coefficient (SCC, green), normalized root-mean-square error (NRMSE, pink), and symmetric mean absolute percentage error (SMAPE, orange) for the correlation functions. A black vertical dashed line is included in all plots to indicate the transition in $r$ between the 1-halo and 2-halo regimes. It is seen that $\xi$ features a $3\%$ error, on average, and $\omega$ features a $5\%$ error. Though exhibiting a larger fractional error, $\eta$ predictions are on average strictly within 1$\sigma$ of the true uncertainty. This similarly holds for $\omega$, and $\xi$ exhibits a bias at large $r$, reflecting the higher fractional error. Both $\xi$ and $\omega$ exhibit large SCC values and low NRMSE and SMAPE values across all bins, indicating good performance. For $\eta$, the SCC value at low $r$ (SCC $\geq 0.5$) indicates a strong correlation between \IAEmu predictions and the ground truth. This gradually decreases at the onset of the 2-halo regime, with the NRMSE and SMAPE performance decreasing as well. }
    \label{fig:metrics}
\end{figure*}

We analyze the performance of \IAEmu, first on the held-out (in-distribution) test set, and further on a set of IA observations from the \textsc{tng300} suite of hydrodynamical simulations.
\IAEmu achieves high average accuracy for both galaxy position-position and position-orientation statistics, and demonstrates robustness to shape noise in the orientation-orientation statistics without signs of overfitting.
We show that \IAEmu's performance on $\eta$ is more difficult to quantify due to the high stochasticity of the correlation function, even after averaging over multiple realizations of the data.
We lastly show that when fitting alignment parameters to IA correlations from \textsc{tng300}, \IAEmu has better than $0.4\sigma$ agreement with \halotoolsia. 

\subsection{Performance}
\label{sec:performance}

\textbf{Accuracy}. We evaluate the model on the 20\% in-distribution but held-out test set, as summarized in Figure \ref{fig:metrics}. 
All test-set predictions are mean predictions averaged over 50 forward passes (i.e., predictions with Monte Carlo Dropout) of \IAEmu, so that an epistemic uncertainty on predictions can be retrieved. 
Reported metrics are evaluated on the correlations in their original domain; they are not computed in the standardized or log-transformed domain employed during \IAEmu training. 

In reporting metrics, we exclude outlier examples in the $\omega$ and $\eta$ correlations where \IAEmu shows a strong Spearman correlation coefficient ($>0.5$) with the ground truth or its predicted amplitude is within $1\sigma$ of the true uncertainty of the data for the majority of the bins, but the fractional error exceeds a factor of $100$ and $450$, respectively.
These extreme values arise from the lack of a high-signal-to-noise ground truth and occur when the true correlation amplitude is small and changes sign, conditions under which the fractional error can become arbitrarily large even when predictions accurately follow the underlying signal.  
They are therefore artifacts of the small amplitude and noise of the data, rather than indicators of genuine model inaccuracies.
The thresholds were chosen empirically to exclude only these pathological cases, affecting fewer than $\lesssim 1\%$ of test-set predictions.
Including these outliers does not impact the overall assessment of performance but significantly inflates the mean fractional error, introducing numerical instability.
Notably, the median fractional error remains stable—within approximately $0.1\%$—for both $\omega$ and $\eta$, irrespective of whether these extreme cases are retained.
Even with this mitigation, many instances remain in the test set where the performance of \IAEmu is visually suitable, but features large fractional error due to these numerical artifacts.

For the position-position ($\xi$) correlation, the mean fractional error per bin (top panel) reaches a maximum of $10\%$, with \IAEmu achieving an average error of $3.2\%$ for $\xi$.
The $\xi$ performance is biased high at large $r$, where the $\xi$ correlation amplitudes are small ($||\xi|| \ll 1$) and approach zero.
This bias may arise in part from the standardization for training, which can disproportionately emphasize regions with larger amplitudes or compress the dynamic range at small values, leading to systematic bias.
Additionally, as \IAEmu naturally predicts $\log \xi$, small residuals near zero can appear large when transformed back to linear space.

For $\omega$, the accuracy drops at the onset of the 2-halo regime, with an average model error across all bins of $4.9\%$ and a similar maximum of $\sim10\%$.
We also find that the median fractional error across all bins is less than $10\%$ for $66\%$ of test-set predictions. 
This is approaching the accuracy for IA modeling likely required for Stage IV surveys \citep{Paopiamsap_2024}. 
The mean fractional error for orientation-orientation ($\eta$) is significantly higher, averaging $54\%$. 
However, it is important to note that the ground truth $\omega$ and $\eta$ correlations -- even after averaging over 10 realizations of the dataset -- are generally noisy and can often fluctuate between positive and negative values. 
Fractional error can thus be misleading in this case due to the absence of high-signal ground truth values for comparison, and due to the correlation amplitudes being close to zero and frequently changing sign.
We also studied the $r$-weighted mean fractional errors ((5.5\%, 5.2\%, and 67.4\%) compared to their unweighted counterparts  (3.2\%, 4.9\%, and 53.8\%), which reveals a $\sim2\%$ difference for $\xi$, a similar performance for $\omega$, and a larger difference for $\eta$.
This is in line with the observed bias for $\xi$ at large $r$.
In the case of $\eta$, we emphasize that fractional error for $\eta$ should not be interpreted in isolation as a gauge of model accuracy.

With this in mind, we show in the middle panel of Figure \ref{fig:metrics} the median residual in units of the dataset's true aleatoric uncertainty $\hat{\sigma}$.
From this metric, it is observed that despite the large fractional error in \(\eta\), the predictions of \IAEmu remain strictly within \(1\sigma\) of the ground truth correlations across all bins.
This trend also holds for \(\omega\).
For $\xi$, the residual is computed in log space in the 2-halo regime to more consistently represent the bias with how \IAEmu was trained, and to avoid exaggerated deviations caused by exponentiating small correlation values.
Despite the large stochasticity of $\omega$ and $\eta$, this indicates that \IAEmu has learned to capture the mean behavior and not overfit to the noise fluctuations in these correlations.
This provides the added benefit of capturing the ``cosmic mean'' of the correlations directly with \IAEmu, which would otherwise require running multiple realizations of the underlying HOD.
This can also be frequently seen in example \IAEmu predictions for $\omega$ and $\eta$ as shown in Appendix \ref{sec:appendix-plots}.

\textbf{Metrics.}  We further evaluate the performance of \IAEmu using three key metrics: the Spearman correlation coefficient (SCC), which measures the rank correlation between predicted and true values; the normalized root mean squared error (NRMSE); and the symmetric mean absolute percentage error (SMAPE).
The SCC, which ranges between 0 and 1, is particularly useful for assessing rank-based correlations and is well-suited for analyzing sequence data.
The NRMSE is defined as:
\begin{equation}
\text{NRMSE} = \sqrt{\frac{\frac{1}{n} \sum_{i=1}^{n} (y_i - \hat{y}_i)^2}{\frac{1}{n} \sum_{i=1}^{n} y_i^2}}
\end{equation}
where \( y_i \) represents the ground truth value, and \( \hat{y}_i \) denotes the corresponding prediction by \IAEmu.
This metric provides an indication of prediction accuracy, but can be sensitive to outliers, due to dependence on squared error.
To quantify relative percentage error, we use the SMAPE, which is defined
\begin{equation}
\text{SMAPE} = \frac{1}{n} \sum_{i=1}^{n} \frac{2 |y_i - \hat{y}_i|}{|y_i| + |\hat{y}_i| + \epsilon}
\end{equation}
where \( \epsilon = 10^{-8} \) is introduced to prevent division by zero.
The SMAPE is generally more robust to outliers compared to the NRMSE, but it tends to be more sensitive to small values (i.e., small correlation amplitudes).
These three metrics are selected due to their scale-invariant properties, which are essential for comparing \IAEmu's performance across the varying scales of \( \xi \), \( \omega \), and \( \eta \).
An SCC value of 1 indicates a perfect correlation between \IAEmu and the ground truth data, while lower values of NRMSE and SMAPE reflect better predictive performance.
The reported metrics are averaged over correlations in the test set.
Together, these metrics provide a comprehensive assessment of \IAEmu's performance.

For $\xi$, we find an SCC value of $0.99$ when averaged across all bins, and a value of $0.98$ for $\omega$ as seen in Figure \ref{fig:metrics}.
This indicates a very strong correlation between the \IAEmu predictions and the underlying data.
For $\eta$, the average SCC across all bins is $0.55$, with the $\text{SCC} \approx 0.75$ at low $r$, but it is around 0.5 after entering the 2-halo regime, which still reflects a moderate correlation between the data and model.
At larger $r$, the SCC decreases, indicating a weak correlation. It is important to note that the SCC can be strongly affected by stochasticity and the low amplitude of the data, particularly when the amplitudes approach zero, as is the case frequently for $\eta(r)$.

For $\xi$ and $\omega$, the NRMSE averaged across all bins is $0.19$, as shown in the bottom panel of Figure \ref{fig:metrics}.
The corresponding SMAPE values averaged across all bins are $\text{SMAPE}(\xi) = 0.10$ and $\text{SMAPE}(\omega) = 0.30$.
The relatively low NRMSE values indicate that, on average, the predictions closely follow the ground truth across the full range of data.
However, the higher SMAPE for $\omega$ compared to $\xi$ suggests that the relative error is more pronounced for $\omega$, potentially due to the generally smaller correlation amplitudes in $\omega$.
This implies that while the absolute prediction error remains comparable, the percentage error is exacerbated by the lower magnitude of the true values in $\omega$.
A similar trend is observed for $\eta$, where both the NRMSE ($0.69$) and SMAPE ($1.11$) are significantly larger.
These higher values indicate that \IAEmu's predictions for $\eta$ exhibit larger absolute and relative deviations from the ground truth.
This could be attributed to a decrease in performance, increased variability, or a broader dynamic range in $\eta$, which naturally poses greater challenges for accurate predictions.

\textbf{Limitations}. \IAEmu's predictions for $\eta$ are less accurate compared to $\xi$ and $\omega$, which perform well across metrics considered.
While \IAEmu successfully captures the correct scaling of $\eta$ across all bins, its accuracy for $\omega$ and $\eta$ is primarily limited by the stochastic nature of these correlations, even when trained on multiple realizations and evaluated on their means.
As demonstrated by examples in Appendix \ref{sec:appendix-plots}, the averaged ground truth correlations still exhibit fluctuations that are indicative of noise due to the relatively small volume considered for the simulations.
This hinders the evaluation of \IAEmu's performance as well as training; however, as also demonstrated in the middle panel of Figure \ref{fig:metrics}, \IAEmu reliably captures the underlying mean behavior despite the presence of noise.
The bias at large $r$ for $\xi$ can likely be attributed to the use of $\log$ and the standardization procedure for training.
In a multitask framework, standardization can potentially be avoided by using trainable loss coefficients \citep{kendall2018multitasklearningusinguncertainty}.

\textbf{Efficiency}. We emphasize the stark difference in speed for obtaining correlations given input HOD parameters using \IAEmu versus \texttt{\halotoolsia}.
\IAEmu performs inference on a batch of size $32,768$ in 1.02 seconds on a single NVIDIA A100-80GB GPU, while the HOD, when run in parallel on 150 CPU cores for the same parameters, takes approximately 3 hours.
This constitutes an approximate factor of $10^4$ improvement in runtime.
On a single CPU core, this would constitute an improvement of roughly $10^6$.
While a direct comparison between a GPU and multiple CPU cores is inherently challenging due to differences in hardware architectures and parallelization capabilities, this comparison highlights the practical advantage of \IAEmu in terms of computational efficiency for large-scale inference tasks with typical hardware availability.
Additionally, \IAEmu's compatibility with differentiable sampling algorithms allows for rapid posterior estimation, further showcasing its efficiency in inverse modeling applications.

\subsection{Aleatoric and Epistemic Uncertainty in \IAEmu Predictions}
\label{sec:uncertainty}

\begin{figure*}
    \centering
    \begin{minipage}[t]{0.48\textwidth}
        \centering
        \includegraphics[width=\linewidth]{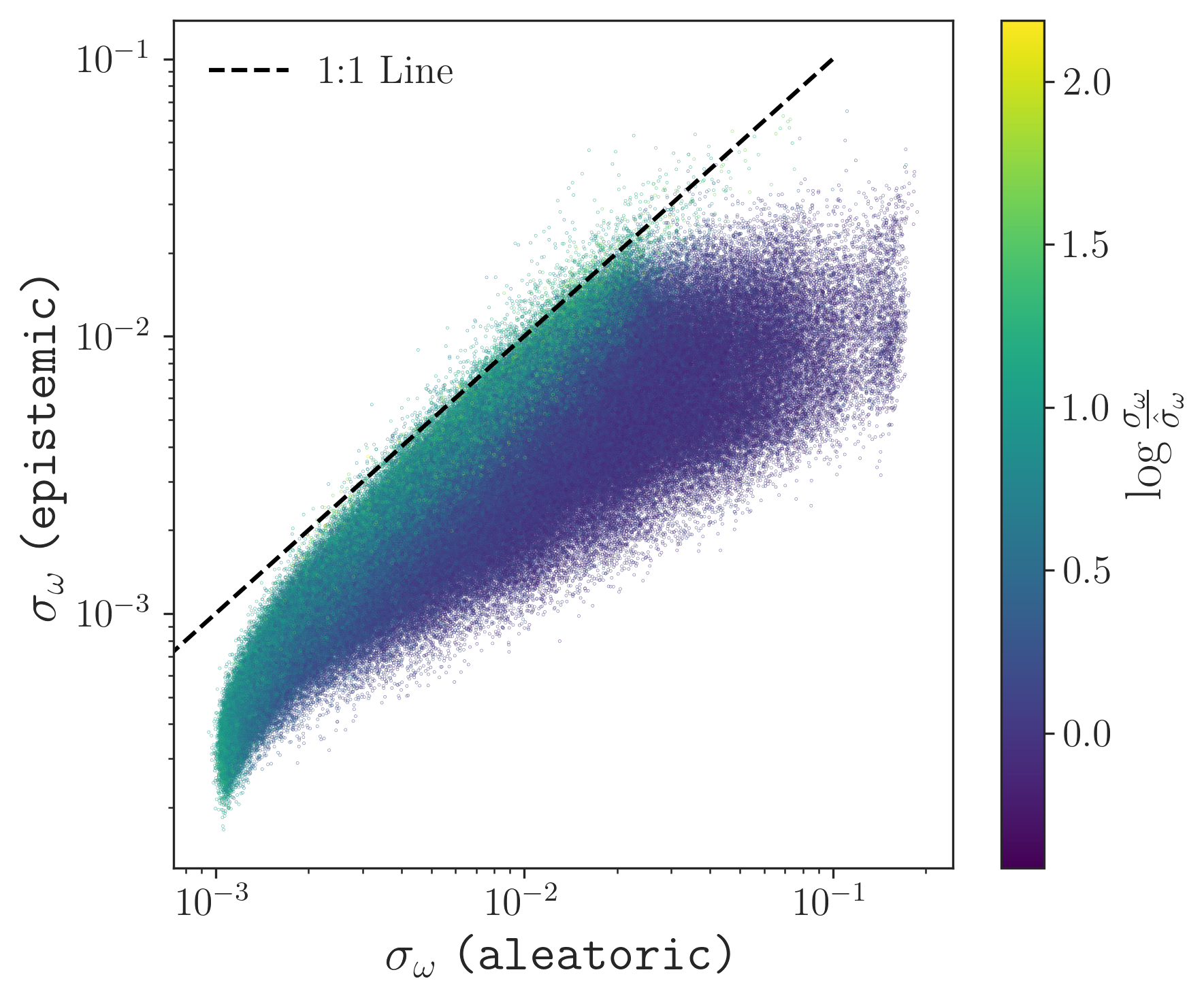}
    \end{minipage}\hfill
    \begin{minipage}[t]{0.48\textwidth}
        \centering
        \includegraphics[width=\linewidth]{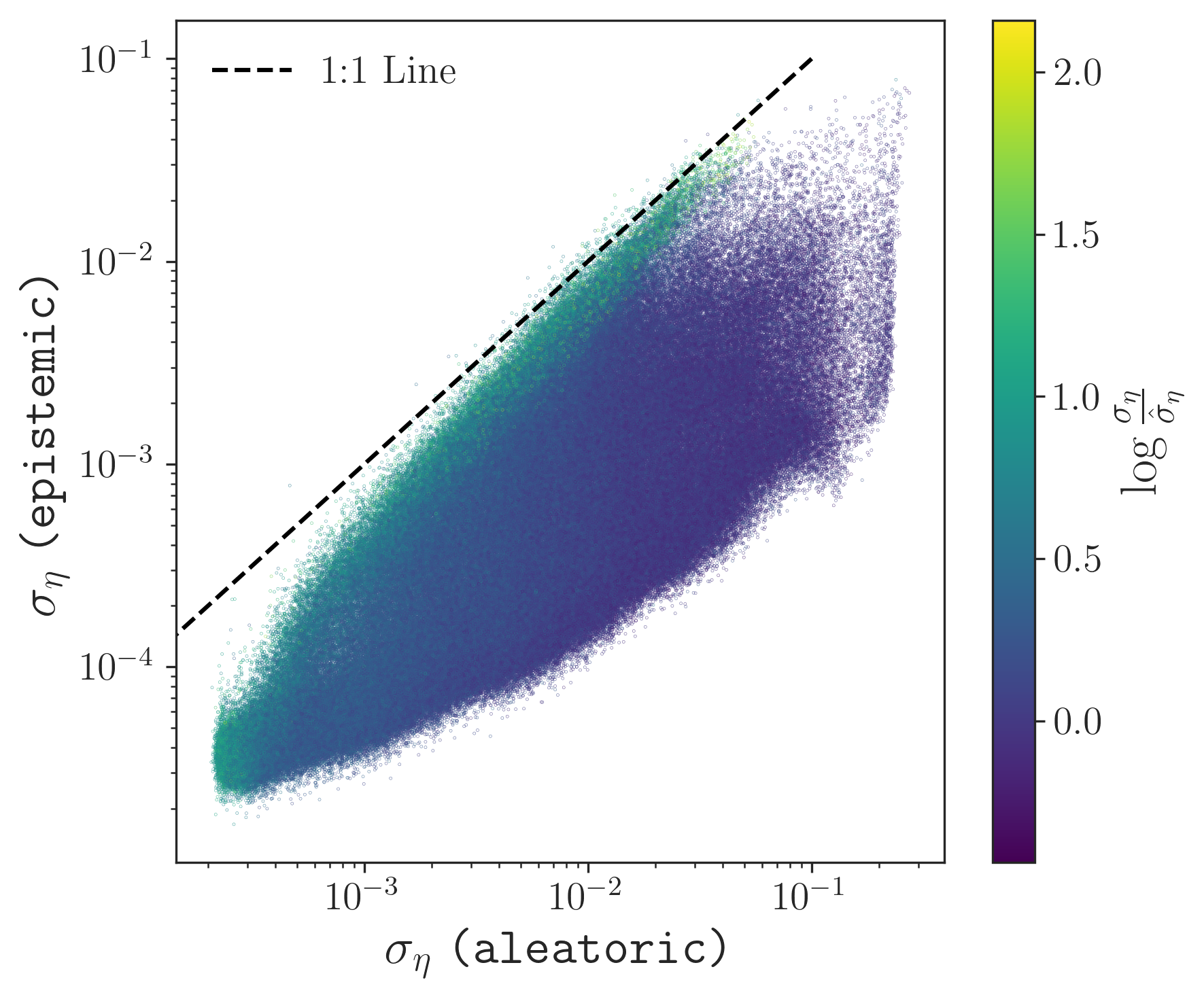}
    \end{minipage}
    \caption{Aleatoric vs. epistemic uncertainty comparison for $\omega$ and $\eta$ with uncertainty bias.
    For test-set predictions, we analyze the total spread of aleatoric uncertainties of the data predicted by \IAEmu and epistemic uncertainties due to the stochasticity of \IAEmu. The coloring corresponds to the log-residual between \IAEmu predicted aleatoric uncertainties and (true) aleatoric uncertainties from \halotoolsia produced from the 10 realizations used in producing the dataset. It is seen that the epistemic uncertainty is generally smaller than the aleatoric uncertainty, due to the majority of the scatter falling below the 1:1 line in aleatoric-epistemic uncertainty space. A general bias of 0.42 dex for $\omega$ and 0.24 dex for $\eta$ is observed between the true and predicted aleatoric uncertainties, with \IAEmu uncertainty estimates being biased high. This is exacerbated near the 1:1 line, in which the epistemic uncertainty of \IAEmu is comparable to the predicted aleatoric uncertainty.}
    \label{fig:coverage_plots}
\end{figure*}

\begin{figure*}
    \centering
    \begin{minipage}[t]{0.48\textwidth}
        \centering
        \includegraphics[width=\linewidth]{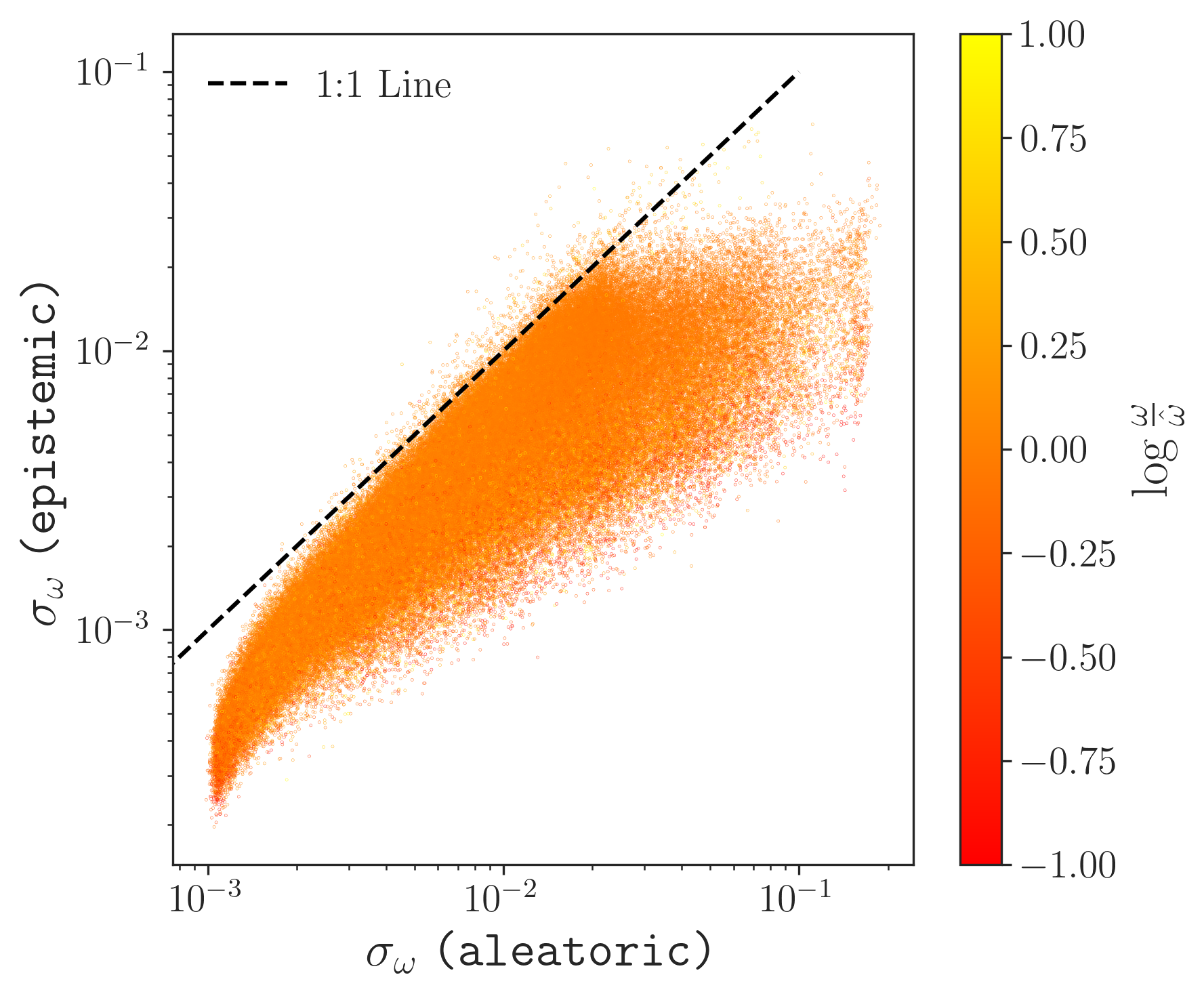}
    \end{minipage}\hfill
    \begin{minipage}[t]{0.48\textwidth}
        \centering
        \includegraphics[width=\linewidth]{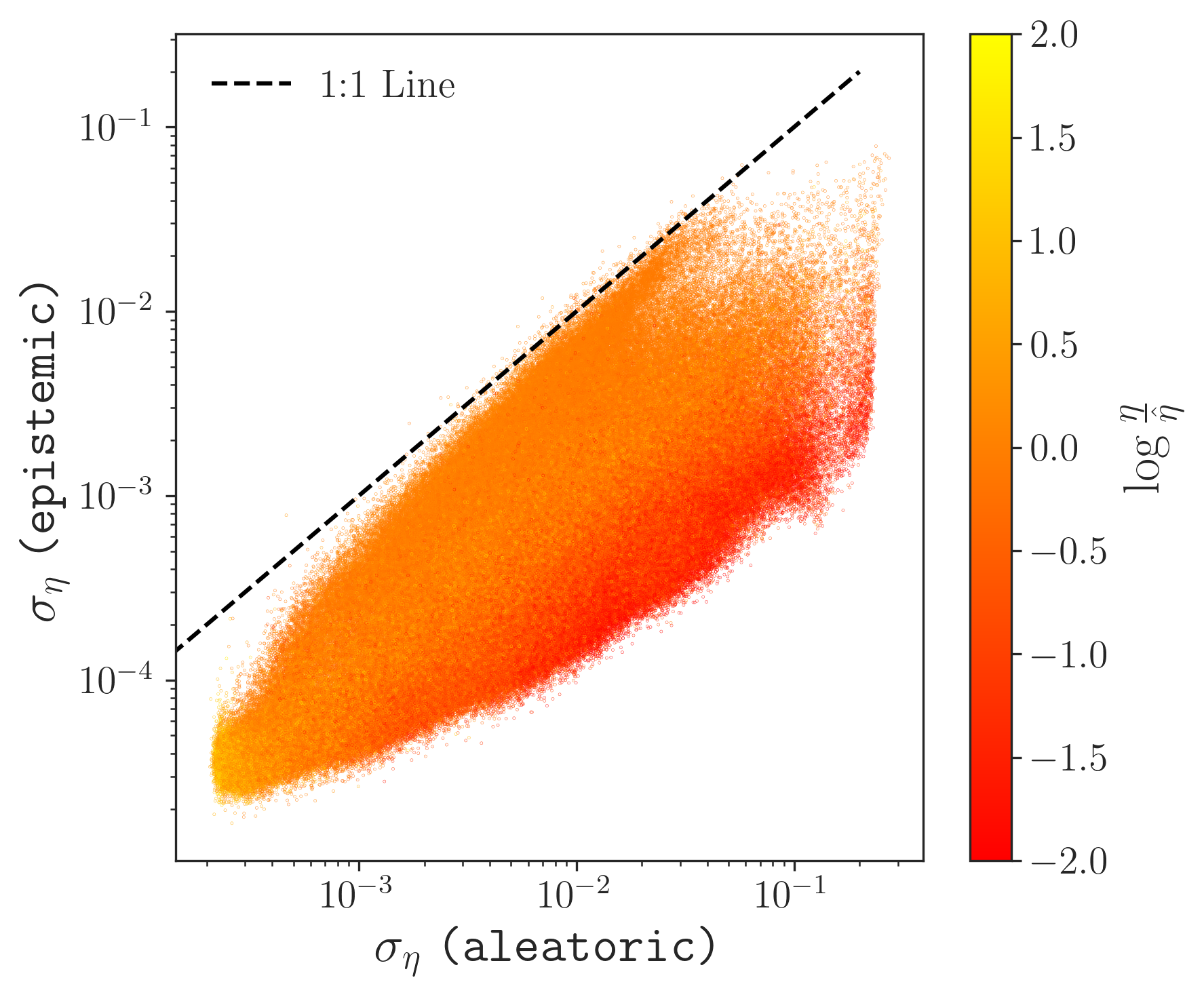}
    \end{minipage}
    \caption{Aleatoric vs. epistemic uncertainty comparison for $\omega$ and $\eta$ with correlation amplitude bias. For test-set predictions, we analyze the total spread of aleatoric uncertainties of the data predicted by \IAEmu and epistemic uncertainties due to the stochasticity of \IAEmu. The coloring corresponds to the log-residual between \IAEmu predicted correlation amplitudes and (mean) ground truth amplitudes from \halotoolsia produced from the 10 realizations used in producing the dataset. It is seen that there is no clear correlations between residuals in the amplitudes and \IAEmu aleatoric and epistemic uncertainties in the case of $\omega$. For $\eta$, it is seen that the sharpest log-residual occurs for predictions in the region where the \IAEmu aleatoric uncertainty is $\approx 2$ dex larger than the associated epistemic uncertainties. This can be an instance of \IAEmu overfitting, wherein the intrinsic uncertainty of the model on the correlation amplitude is negligible compared to the correlations own uncertainty.}
    \label{fig:coverage_plots2}
\end{figure*}

Due to the high stochasticity of correlations like $\omega$ and $\eta$, \IAEmu was designed to produce \emph{distributions} on its outputs, tracking multiple types of uncertainty, thereby enabling confidence assessment in its predictions.

Aleatoric uncertainty represents the intrinsic variability in the data, in this case representing variance in the correlations due to galaxy shape noise and sample variance, as studied in \citet{vanalfen_2023}.
The aleatoric uncertainties of $\omega$ and $\eta$ can thus be reduced through a larger simulation box size (resulting in more galaxies) and through multiple realizations of the same volume.
Shape noise dominates over sample variance in the HOD model predictions \citep{vanalfen_2023}, making multiple realizations important for retrieving accurate correlation functions.
Epistemic uncertainties are uncertainties inherent to a model and can be large when an architecture is ill-suited for a task, or when a model is not trained on sufficient data \citep{Hullermeier_2021}.
Aleatoric uncertainties are directly output from \IAEmu through its design and training procedure.
Epistemic uncertainties are obtained via the Monte Carlo dropout technique \citep{gal2016dropout}, where dropout is used during inference across multiple forward passes.
This introduces stochasticity into \IAEmu's predictions, and the resulting variance in the outputs represents the epistemic uncertainty (see \citet{Hullermeier_2021} for a review on distinguishing between aleatoric and epistemic uncertainty).

Figure \ref{fig:coverage_plots} compares the aleatoric and epistemic uncertainties from \IAEmu with the true aleatoric uncertainty from \halotoolsia across 10 realizations of the simulation.
The figure shows that epistemic uncertainties are generally smaller than aleatoric uncertainties, as indicated by the majority of scatter points falling below the 1:1 line.
This suggests that \IAEmu's architecture is sufficiently expressive for this task, and that it was not data-limited during training despite the stochasticity of these correlations.
However, a median bias of 0.34 dex for $\omega$ and 0.18 dex for $\eta$ for aleatoric uncertainties when compared to the true aleatoric uncertainties is observed, suggesting that \IAEmu is not perfectly calibrated for aleatoric uncertainties.
This residual is particularly pronounced near the 1:1 line, wherein \IAEmu's epistemic uncertainty predictions are comparable to the aleatoric uncertainty predictions.
That is, \IAEmu tends to overestimate aleatoric uncertainties in regions where the correlation amplitudes are less certain.
Nevertheless, the shape noise estimates from \IAEmu can provide valuable covariance information for Monte Carlo inference \citep{berman2025softclustering}, significantly improving posterior constraints compared to inference without covariance information.

We also study the relationship between aleatoric and epistemic uncertainties in terms of the residuals in the correlation amplitudes, as shown in Figure \ref{fig:coverage_plots2}.
For $\omega$, we observe a trend where the largest errors in the correlation amplitudes occur in the regime where the epistemic uncertainties are 1 dex smaller than the predicted aleatoric uncertainties.
This trend is more pronounced for $\eta$, where the highest errors occur when the aleatoric uncertainties are 2 dex larger than \IAEmu's epistemic uncertainties, as seen in Figure \ref{fig:coverage_plots}.
This may indicate an overconfidence for \IAEmu predictions of $\eta$ in this regime; however, it is also clear in comparing Figure \ref{fig:coverage_plots} and Figure \ref{fig:coverage_plots2} that this regime is where the $\eta$ correlations are the noisiest.
In other words, this occurs when the galaxy shape noise is most prominent.
It is thus expected that the residual on \IAEmu predictions would be exaggerated due to \IAEmu not overfitting to the shape noise.
Nonetheless, this regime is also where \IAEmu aleatoric uncertainty predictions are the most accurate.

These insights lead us to the following conclusions about the performance of \IAEmu for $\omega$ and $\eta$, and provide a useful diagnostic for gauging its accuracy in the absence of \halotoolsia ground truth data: 
\begin{itemize} 
\item \IAEmu is not significantly limited by data, as evidenced by the scale of its epistemic uncertainties compared to aleatoric uncertainties for $\omega$ and $\eta$. 
\item  \IAEmu residuals on correlation amplitudes are largest when both the true and predicted aleatoric uncertainties of IA correlations are large, which is typically an artifact of \IAEmu learning the mean behavior of these noisier statistics.
\item \IAEmu tends to overestimate aleatoric uncertainties for both $\omega$ and $\eta$ in regimes where they are comparable to the epistemic uncertainties. This occurs when the model is most uncertain. \IAEmu correlation amplitudes are still accurate in this regime, as shown in Figure \ref{fig:coverage_plots2}.
\item \IAEmu aleatoric uncertainty predictions are most accurate for regions of parameter space that yield the noisiest correlations. This can be attributed to stronger gradient information with larger variance magnitudes, as seen in the Equation \ref{eq:NLL}.
\end{itemize}

In practice, one may consider both the aleatoric and epistemic uncertainties predicted by \IAEmu to assess the quality of its predictions in the absence of an underlying \texttt{\halotoolsia} ground truth.
Despite the observed bias, aleatoric uncertainty remains valuable for covariance estimation (e.g., accounting for shape noise) when performing parameter inference with \IAEmu \citep{berman2025softclustering}.
Post-hoc calibration methods, such as those discussed in \citet{Grand_n_2022}, can help correct for these biases in parameter inference.
These methods calibrate uncertainty estimates after training, ranging from non-parametric approaches like histogram binning and isotonic regression, to simple parametric schemes such as Platt scaling \citep{guo2017calibrationmodernneuralnetworks} and its extensions (e.g.\ temperature scaling and Dirichlet calibration).
Even when the primary concern is the correlation amplitudes, the relationship between \IAEmu's epistemic and aleatoric uncertainties provides valuable insight into the reliability of the predictions, as illustrated in Figures \ref{fig:coverage_plots} and \ref{fig:coverage_plots2}.

\subsection{Parameter Estimation with \textsc{tng300}}
\label{sec:illustris}

\begin{figure*}
    \centering
     \includegraphics[width=\textwidth]{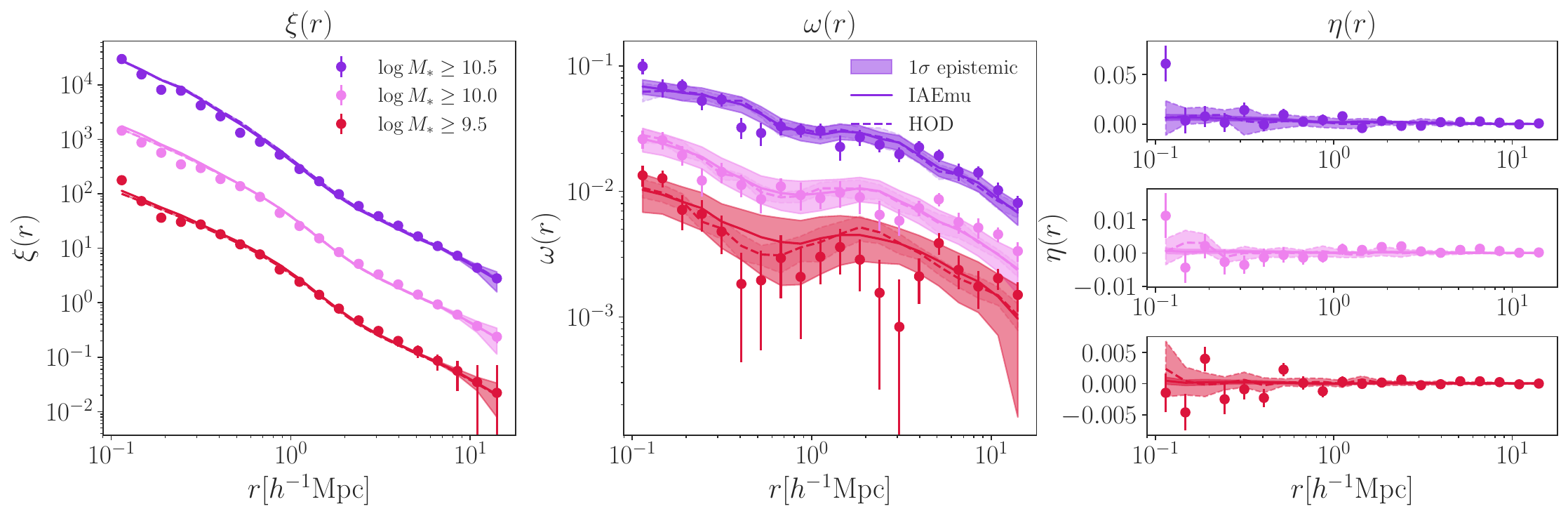}
    \caption{The two-point correlation functions (2PCFs) for IA, fitted to observations from the \textsc{tng300} simulation, using both \halotoolsia and \IAEmu. The correlations are measured across three mass threshold samples, as denoted in the left panel legend. Purple corresponds to most massive sample, pink for intermediate, and red for least massive. True correlations are shown as scatter points and HOD and \IAEmu fits shown as lines. These 2PCFs correspond to the posterior mean values of $\mu_{\text{cen}}$ and $\mu_{\text{sat}}$, as shown in Figure \ref{fig:corner}. Error bars for \textsc{tng300} are obtained via jackknife resampling, while the $1\sigma$ epistemic uncertainty for \IAEmu is estimated from 50 forward passes using the Monte Carlo dropout technique. The $1\sigma$ uncertainty band for \halotoolsia reflects variations from random realizations of the model. \textbf{Left:} Position-position correlation function $\xi$ with the upper and lower curves offset by 1 dex for visual clarity, showing that \IAEmu can model galaxy bias. 
    \textbf{Middle:} Position-orientation correlation function $\omega$. \textbf{Right:} Orientation-orientation correlation function $\eta$.}
    \label{fig:iaemuhodcomparison}
\end{figure*}

Our results in the previous section were obtained using a test set held out during training from the HOD simulation dataset.
This demonstrates that the model generalizes well to novel data drawn from the same distribution as the training set.
Previously, \citet{vanalfen_2023} showed that \texttt{\halotoolsia} is expressive enough to model the IA signal derived from The \textsc{tng300} suite of hydrodynamical simulations, which incorporate more complex physics, including baryonic effects.
This constitutes an out-of-distribution (OOD) shift over the joint distribution of inputs and outputs from an HOD that \IAEmu was trained on.
In this section, we investigate whether \IAEmu exhibits a similar modeling capability as \halotoolsia, and can thus be robust to OOD shifts for inverse modeling.
To this end, we select the best-fit occupation model parameters that reproduce the HOD of \textsc{tng300}, as described in \citet{vanalfen_2023}, and determine the posterior distributions on $\mu_\text{cen}$ and $\mu_\text{sat}$ that fit the signal.
This ensures that halos with comparable masses are populated with a comparable number of galaxies as in \textsc{tng300}, leaving galaxy alignment as the major factor affecting how similar correlations from the two samples are.
This experiment therefore enables us to investigate potential biases between \IAEmu and \texttt{\halotoolsia} in the alignment parameter input space when modeling IA for an OOD sample.

To perform parameter inference, we leverage the differentiability of \IAEmu to attain efficient posterior estimates.
One of the advantages of \IAEmu, and neural networks in general, is its ability to act as a differentiable forward model.
This property is particularly useful for inverse problems, where the goal is to perform parameter inference based on given observations.
By exploiting this differentiability, one can employ a range of differentiable sampling algorithms to obtain posterior distributions for the parameters.
While \citet{vanalfen_2023} used MCMC for this purpose, we instead use Hamiltonian Monte Carlo (HMC) \citep{DUANE1987216} to achieve posterior estimates more efficiently by leveraging the gradient information inherent in \IAEmu.
Further theoretical background for HMC can be found in Appendix \ref{sec:appendix-training}.

We employ two different methods: \IAEmu\ with HMC and \texttt{\halotoolsia} with MCMC.
In both cases, we model $\omega$ by applying a uniform prior on the alignment parameters: $\mu_{\text{cen}}, \mu_{\text{sat}} \sim U(-1,1)$.
We fit to $\omega$ because $\xi$ lacks any dependence on the IA parameters, as it represents the galaxy clustering.
In contrast, $\eta$ does incorporate IA information; however, the noise associated with this statistic presents a much greater challenge for solving the inverse problem compared to $\omega$.
Both the HMC and MCMC experiments employed the same jackknife covariance matrix, estimated from the \textsc{tng300} data itself.

We employ Hamiltonian Monte Carlo (HMC) with the No U-Turn Sampler (NUTS, \cite{hoffman2011nouturnsampleradaptivelysetting}), using $2000$ warm-up steps and an initial learning rate of $0.005$, collecting $4000$ posterior samples for analysis.
All posteriors resulted in an effective sample size greater than 1000, and all HMC experiments were executed on a single GPU and converged in roughly one minute.
For comparison, the MCMC implementation in \citet{vanalfen_2023} utilized parallelization across 150 CPU cores and required up to a full day due to computational constraints.
This highlights a near $2000\times$ speed-up for \IAEmu-HMC relative to \halotoolsia-MCMC on the tested hardware.
While this is somewhat lower than the acceleration achieved in forward modeling with \IAEmu compared to \halotoolsia, it remains a substantial improvement.
The reduced gain is anticipated: HMC necessitates backpropagation through \IAEmu, roughly doubling the computational load compared to a forward pass \citep{Goodfellow-et-al-2016}, along with added overhead from NUTS numerical integration.
Additionally, the sequential nature of HMC does not allow for parallelization.
Nevertheless, its rapid convergence demonstrates its efficiency over traditional, parallelized MCMC approaches.
We emphasize that a detailed convergence comparison was not performed, and that the parallelized MCMC yielded approximately $75,000$ posterior samples, in contrast to the $4000$ obtained via \IAEmu.
Hence, the reported speed-up metrics should be interpreted as indicative benchmarks rather than definitive measurements.

\begin{figure*}
    \centering
    \subfigure[$\log M_* \geq 10.5$]{
        \includegraphics[width=0.31\textwidth]{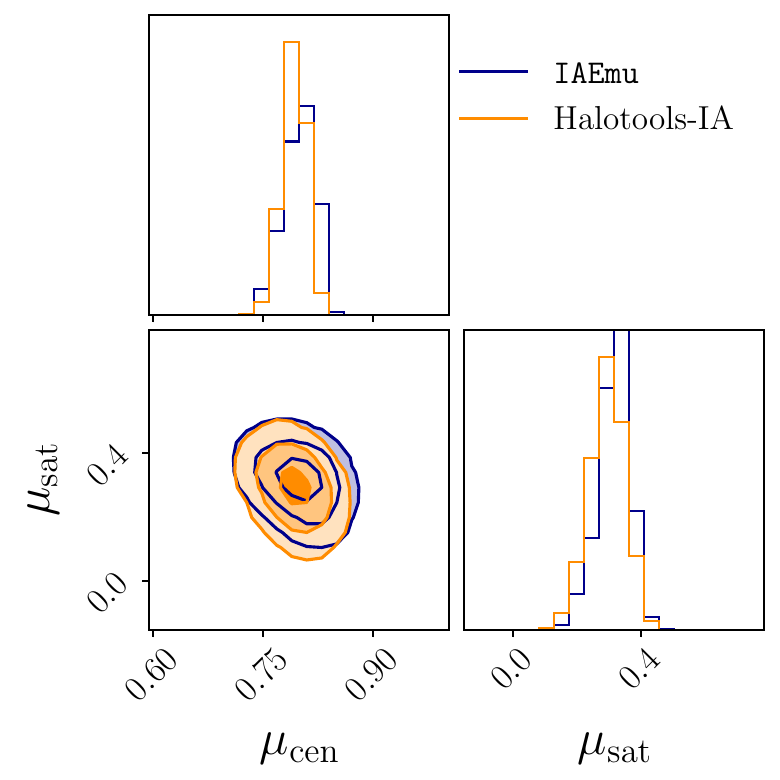}
        \label{fig:figure1}
    }
    \subfigure[$\log M_* \geq 10.0$]{
        \includegraphics[width=0.31\textwidth]{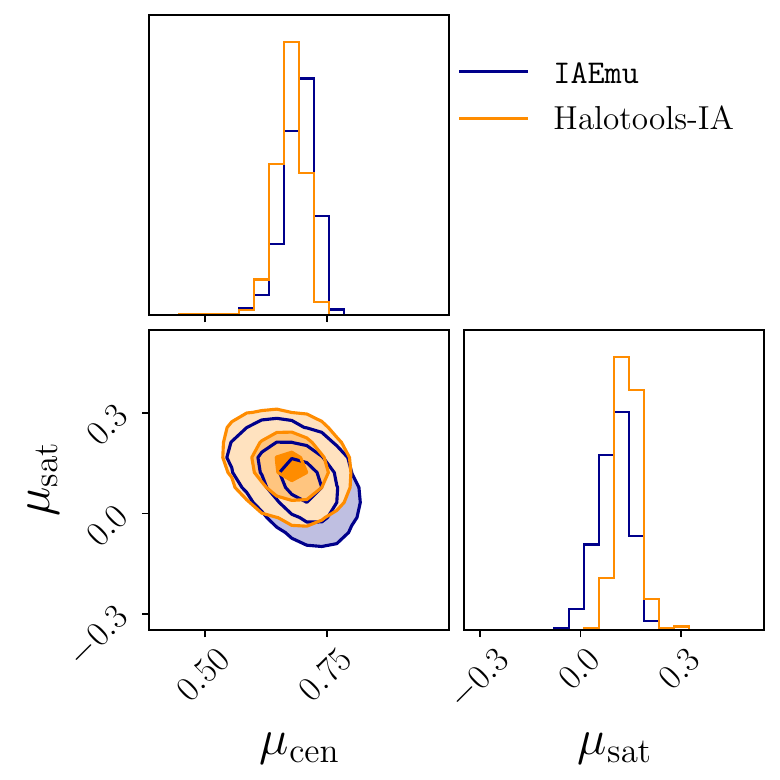}
        \label{fig:figure2}
    }
    \subfigure[$\log M_* \geq 9.5$]{
        \includegraphics[width=0.31\textwidth]{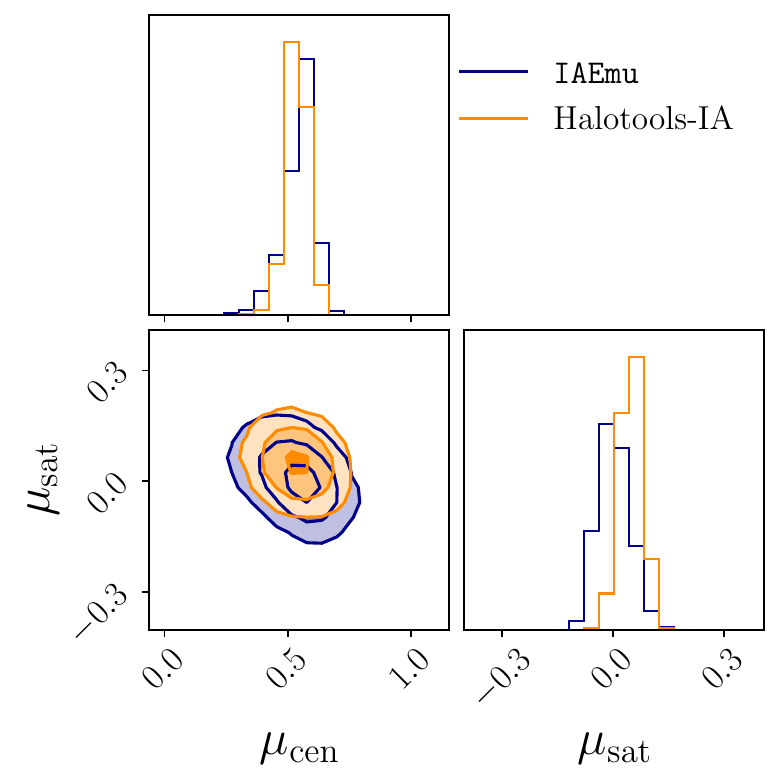}
        \label{fig:figure3}
    }
    \caption{
    Optimal parameter values for central alignment strength ($\mu_{\text{cen}}$) and satellite alignment strength ($\mu_{\text{sat}}$) fit to $\omega$ observations from \textsc{tng300} with three distinct mass cutoffs for halos included in the underlying HOD model. Posterior contours for \halotoolsia and \IAEmu are shown with 4000 posterior samples each, with \halotoolsia contours in orange and \IAEmu contours in blue. Posteriors for \halotoolsia were obtained via MCMC using 75 walkers running in parallel for 23 hours on CPU, resulting in up to 1300 steps per walker, or as few as about 450 steps per walker for slower runs. Posteriors for \IAEmu were retrieved using NUTS, a variant of the HMC algorithm, with 2000 warm-up steps around a minute on a single GPU. $\IAEmu$ posteriors exhibit a better than $0.4 \sigma$ overlap with posteriors from \halotoolsia, indicating that \IAEmu can generalize to OOD shifts for inverse modeling. Exact posterior summaries for comparison can be found in Table \ref{tab:table}. \textbf{Left:} Sample 1 \IAEmu posteriors with optimal values $\mu_{\text{cen}} = 0.81$ and $\mu_{\text{sat}} = 0.35$. \textbf{Middle:} Sample 2 \IAEmu posteriors with optimal values $\mu_{\text{cen}} = 0.70$ and $\mu_{\text{sat}} = 0.14$. \textbf{Right:} Sample 3 \IAEmu posteriors with optimal values $\mu_{\text{cen}} = 0.52$ and $\mu_{\text{sat}} = 0.01$.}
    \label{fig:corner}
\end{figure*}

The correlation function predictions from \IAEmu with posterior means for $\mu_\text{cen}$ and $\mu_\text{sat}$ are shown in Figure \ref{fig:iaemuhodcomparison}, in which we see that there is generally good agreement with \IAEmu predictions compared to those from \halotoolsia for all correlations.
The correlation function, $\xi$, is also shown to illustrate the agreement between \halotoolsia and \IAEmu for galaxy clustering statistics; however, it does not depend on $\mu_\text{cen}$ or $\mu_\text{sat}$.
We also see that the quality of fit for both \halotoolsia and \IAEmu decreases with decreasing stellar mass, which was also observed in \citet{vanalfen_2023}.
This provides some indication that a constant alignment strength parameterization is not sufficient to extend this HOD-based model to hydrodynamic simulations in the low-mass regime, which could potentially be addressed with a mass-dependent alignment parameterization.

The corner plots in Figure \ref{fig:corner} show the joint $(\mu_{\text{cen}},\mu_{\text{sat}})$ posteriors for three separate stellar mass thresholds $M_*$ for both \IAEmu and \halotoolsia.
Sample 1 corresponds to $\log (M_*) > 10.5$, Sample 2 to $\log (M_*) > 10.0$, and Sample 3 to $\log (M_*) > 9.5$.
The HOD parameter fits corresponding to these mass cutoffs can be found in \citep{vanalfen_2023}.
We confirm two trends also observed in \citet{vanalfen_2023}: central alignment strength is larger than satellite alignment strength, and the alignment strength monotonically increases with the stellar mass threshold.
We find a greater than $0.4\sigma$ agreement between MCMC with \halotoolsia and HMC with \IAEmu for all samples.
The strongest discrepancy is in the posterior variance for Sample 3, which is the noisiest set of \textsc{tng300} data and also has the largest \IAEmu epistemic uncertainty, as seen in Figure \ref{fig:iaemuhodcomparison}.
This reflects the discussion in Section \ref{sec:uncertainty}, wherein it was seen that the epistemic uncertainty of \IAEmu is correlated with the true and predicted aleatoric uncertainty.
Exact values for \halotoolsia and \IAEmu posterior summary statistics are shown in Table \ref{tab:table}. 

\begin{table}
\centering
\renewcommand{\arraystretch}{1.5}  
\caption{Posterior values for \IAEmu and \halotoolsia fit on \textsc{tng300}.}
\begin{tabular}{|c|c|c|c|c|}
\hline
\textbf{Sample} & \textbf{Mass Cutoff} & \textbf{Posterior} & $\mu_{\text{cen}}$ & $\mu_{\text{sat}}$ \\
\hline
\multirow{2}{*}{1} & log M$_{*} > 10.5$ & \IAEmu & $0.80^{+0.02}_{-0.03}$ & $0.32^{+0.04}_{-0.05}$ \\
                   &                   & \halotoolsia & $0.79^{+0.02}_{-0.02}$ & $0.30^{+0.05}_{-0.06}$ \\
\hline
\multirow{2}{*}{2} & log M$_{*} > 10.0$ & \IAEmu & $0.69^{+0.03}_{-0.03}$ & $0.11^{+0.04}_{-0.05}$ \\
                   &                   & \halotoolsia & $0.68^{+0.03}_{-0.03}$ & $0.14^{+0.03}_{-0.03}$ \\
\hline
\multirow{2}{*}{3} & log M$_{*} > 9.5$ & \IAEmu & $0.56^{+0.04}_{-0.07}$ & $0.00^{+0.05}_{-0.04}$ \\
                   &                   & \halotoolsia & $0.54^{+0.04}_{-0.04}$ & $0.05^{+0.03}_{-0.03}$ \\
\hline
\end{tabular}
\label{tab:table}
\end{table}

\FloatBarrier  

\section{Summary \& Discussion}
In this work, we have developed a neural network-based surrogate model, \IAEmu, designed to predict galaxy intrinsic alignment correlations derived from halo occupation distribution modeling.
\IAEmu eliminates the need to generate full galaxy catalogs and computes correlation functions using traditional HOD pipelines, which are computationally expensive.
On a single GPU, \IAEmu achieves a $\boldsymbol{\times 10^4}$ speed-up in wall-clock time compared to \halotoolsia run on a moderately parallelized CPU setup representative of typical resources (e.g., $\sim$150 cores).
When comparing single GPU to single CPU performance, this corresponds to an approximate $10^6\times$ speed-up.
This substantial acceleration enables efficient forward modeling and significantly expedites inverse modeling tasks.
The differentiable nature of \IAEmu facilitates the use of gradient-based inference methods such as Hamiltonian Monte Carlo (HMC), which are otherwise infeasible with \halotoolsia.
Although the speed-up in individual evaluations is dramatic, the end-to-end improvement in sampling-based inference compared to parallelized MCMC is somewhat lower, due to the additional computational overhead from gradient evaluations and the inherently sequential nature of HMC arising from numerical trajectory integration.
Nevertheless, HMC achieves significantly faster convergence than parallelized MCMC, making it a far more efficient option overall for inverse modeling despite the reduced relative speed-up.

\IAEmu was also designed to account for both aleatoric and epistemic uncertainties, corresponding to the uncertainty inherent in the data and the model, respectively.
This enables confidence assessments for \IAEmu predictions in the absence of ground truth data, as well as provides covariance information for inverse modeling with \IAEmu.
To isolate aleatoric uncertainty, we trained \IAEmu using a mean-variance estimation framework under the assumption of Gaussian-distributed outputs, optimized with the $\beta$-negative-log-likelihood loss function.
For epistemic uncertainty, we employed the Monte Carlo dropout technique, which randomly nullifies certain nodes within \IAEmu during inference, introducing stochasticity into the model predictions.
We find that analyzing these distinct sources of uncertainty provides valuable insight into the strengths and weaknesses of \IAEmu, offering a practical method for diagnosing the quality of emulator predictions and motivating future improvements.

\textbf{Challenges.} 
The \IAEmu architecture and training algorithm were tailored to address the challenging goal of simultaneously modeling galaxy bias and IA correlations from a given set of HOD and alignment parameters.
Architecturally, \IAEmu employs a fully connected embedding network with a 1D convolutional neural network decoder, which features multiple branches.
The shared encoded representation captures the common features of HOD simulations relevant to all correlations, while the separate decoder branches are intended to learn the estimators for each correlation function.
The 1D convolutional layers are crucial for identifying local structures within the encoded representations and for facilitating the transition from input feature vectors to correlation function sequences.
Additionally, \IAEmu incorporates residual connections and dropout layers to improve training convergence and mitigate overfitting.

Simultaneously training on three distinct correlations, each with varying numerical scales and signal-to-noise presented additional engineering challenges from a training perspective.
In particular, the low signal-to-noise of the $\eta$ correlation made accurate modeling of its uncertainties especially difficult.
To address this, we implemented the $\beta$-NLL loss within a multitask learning framework.
This approach ensured that correlations with larger amplitudes did not disproportionately dominate the loss landscape, while facilitating effectively individualized training for each decoder branch, allowing us to prioritize more accurate aleatoric uncertainty estimates for the noisier $\omega$ and $\eta$ correlations while focusing on correlation amplitude predictions for the higher-signal $\xi$ correlation.

\textbf{Results.} \IAEmu achieves an average error of approximately $3\%$ in emulating position-position and $5\%$ in position-orientation galaxy IA correlations.
Although the orientation-orientation correlation $\eta$ is inherently noisier and thus more difficult to quantify performance for, \IAEmu’s predictions for $\eta$ on average remained within $1\sigma$ of the true aleatoric uncertainty of the data when evaluated on the test set.
This indicates that \IAEmu still successfully captures the average behavior of this correlation without overfitting to the shape noise, which would otherwise require multiple realizations of \halotoolsia.
\IAEmu also generally exhibits strong SCC values with the data across all three correlations, indicating that despite the large fractional errors, NRMSE, and SMAPE in the case of $\eta$, \IAEmu captures the overall shape of the correlations well. 

Finally, we found that \IAEmu has comparable performance to \texttt{\halotoolsia} when used to fit the alignment parameters $\mu_{\text{cen}}$ and $\mu_{\text{sat}}$ to IA correlation measurements from the \textsc{tng300} hydrodynamic simulation, in a manner similar to the robustness test originally performed in \citep{vanalfen_2023}.
This demonstrates \IAEmu's robustness to OOD shifts for inverse problems.
Specifically, we observe a better than $0.4\sigma$ agreement in the $\mu_{\text{cen}}$ and $\mu_{\text{sat}}$ posteriors across three separate mass regimes between \IAEmu, fit using HMC, and \halotoolsia, fit with Markov Chain Monte Carlo (MCMC).
A significant advantage was the improvement in computational efficiency; while \halotoolsia with MCMC required approximately one day on a cluster CPU, \IAEmu completed the inverse problem in less than a minute on a single GPU.
This constitutes a nearly $2000\times$ speed up over MCMC with \texttt{\halotoolsia}, demonstrating that the efficiency benefits of neural network surrogate models extend beyond forward modeling. 

\textbf{Limitations.}
\IAEmu is an emulator designed to predict 2PCFs based on HOD modeling.
Consequently, one inherent limitation of \IAEmu is its reliance on phenomenological HOD parameterizations, and \IAEmu thus inherits any limitations currently present in \halotoolsia. 
Moreover, since \IAEmu was trained solely on HODs conducted on the \textsc{Bolshoi-Planck} $N$-body simulation, it does not currently factor in cosmology dependence, which is an avenue of future work.

\textbf{Future Work.} 
There remain several functional improvements required for \IAEmu to be fully deployable in cosmological analyses. 
At present, \IAEmu predicts only 2PCFs and is not configured to model corresponding one-point statistics such as galaxy number counts. 
Furthermore, the 2PCFs computed in \halotoolsia only include galaxy orientations and not their full shapes, which are essential for a complete IA analysis. 
In addition, emerging simulation suites are beginning to forego explicit host–subhalo distinctions in favor of halo-core models \citep{vitório2025exploringcoregalaxyconnection}. 
As a result, \halotoolsia would need to be updated to incorporate halo-core models, and \IAEmu would accordingly need to be retrained. 
Incorporating these improvements is ongoing work in both \halotoolsia and future iterations of \IAEmu.
Lastly, \IAEmu also has the capacity to perform joint inference over both HOD and IA parameters; however, this was not explored in Section~\ref{sec:illustris}. 
The central purpose of this experiment was reproducing the results of \cite{vanalfen_2023}, which focused on varying the IA parameters.
Performing a joint inference over both HOD and IA parameters will be explored in future iterations of \IAEmu.

In future versions of the emulator, we will improve upon the \IAEmu pipeline by exploring different architectural, data, and modeling choices.
\IAEmu, as presented here, operates in a traditional supervised learning regime, where the model learns a direct, deterministic mapping between the HOD parameters and correlations.
Although we introduce stochasticity for uncertainty quantification via MC dropout and MVE, a more natural probabilistic approach could be achieved through conditional diffusion generative modeling, where the model learns a probabilistic mapping via a denoising process on the data, or through flow-based architectures as in \citet{pandey2024charmcreatinghalosautoregressive}.
These models can also have their internal representations restricted by known symmetries in the data, enhancing their effectiveness in physical settings like this \citep[see][for an example of $SO(3)$-equivariant diffusion applied to IA]{jagvaral2023diffusion}, which naturally lends such techniques to field-level modeling of the full galaxy catalog.
Field level emulation can also expand \IAEmu summary statistics outside the 2PCFs modeled here.
The denoising training paradigm when applied to cosmology has thus far exhibited promising results in enhancing the resolution of existing simulations \citep{schanz2023stochasticsuperresolutioncosmologicalsimulations} as well as functioning as surrogate models \citep{mudur2024diffusionhmcparameterinferencediffusion}.

A field-level emulator for \halotoolsia could incorporate elements of NN–based modeling together with differentiable components, such as differentiable HOD models \citep{horowitz2022differentiablestochastichalooccupation}. 
Both NNs and differentiable simulations share a differentiable structure, allowing their components to be integrated within the same modeling framework. 
While non-NN-based differentiable methods provide useful tools for inverse modeling, they can be computationally demanding and are often complemented by faster NN-based surrogates.

Simulation-based modeling has opened a new set of opportunities to better understand galaxy intrinsic alignments, complementing earlier analytic and semi-analytic efforts. 
However, these new techniques have incurred additional computational expense. 
In this work, we have shown a compelling case in which accuracy and efficiency can be achieved with NN-based emulators for galaxy intrinsic alignments from HOD simulations. 
This provides a significant step towards accelerating model validation strategies in preparation for data from the Rubin Observatory, Roman Space Telescope, and other Stage IV surveys.

\section*{Data Availability Statement}
The data underlying this article is publicly accessible and was derived from sources in the public domain, such as the \href{https://www.cosmosim.org}{\textsc{BolshoiPlanck} Catalog} and \href{https://halotools.readthedocs.io/en/latest/}{\texttt{halotools}}.

\section*{Acknowledgements}
We thank Andrew Hearin and Katrin Heitmann for helpful conversations.
S.P. thanks Becky Nevin and Aleksandra Ćiprijanović for useful conversations regarding uncertainty quantification.
S.P. acknowledges support from the National Science Foundation under Cooperative Agreement PHY-2019786 (The NSF AI Institute for Artificial Intelligence and Fundamental Interactions, \url{https://iaifi.org}).
Y.Y. acknowledges support from the Khoury Apprenticeship program. J.B., N.V.A. and Y.Y. are supported in this work by NSF award AST-2206563, the US DOE under grant DE-SC0024787, and the Roman Research and Support Participation program under NASA grant 80NSSC24K0088.
R.W. is supported by NSF award DMS-2134178.
Data generation was conducted on the Discovery cluster, supported by Northeastern University’s Research Computing team.
The machine learning computations were run on the FASRC cluster supported by the FAS Division of Science Research Computing Group at Harvard University.

\newpage
\bibliographystyle{mnras}
\bibliography{bib-3}

@misc{guo2017calibrationmodernneuralnetworks,
      title={On Calibration of Modern Neural Networks}, 
      author={Chuan Guo and Geoff Pleiss and Yu Sun and Kilian Q. Weinberger},
      year={2017},
      eprint={1706.04599},
      archivePrefix={arXiv},
      primaryClass={cs.LG},
      url={https://arxiv.org/abs/1706.04599}, 
}

@article{VanAlfen2025, doi = {10.21105/joss.07421}, url = {https://doi.org/10.21105/joss.07421}, year = {2025}, publisher = {The Open Journal}, volume = {10}, number = {107}, pages = {7421}, author = {Van Alfen, Nicholas and Campbell, Duncan and Hearin, Andrew and Blazek, Jonathan}, title = {Halotools: A New Release Adding Intrinsic Alignments to Halo-Based Methods}, journal = {Journal of Open Source Software}}

@article{Chisari_2013,
   title={Cosmological information in the intrinsic alignments of luminous red galaxies},
   volume={2013},
   ISSN={1475-7516},
   url={http://dx.doi.org/10.1088/1475-7516/2013/12/029},
   DOI={10.1088/1475-7516/2013/12/029},
   number={12},
   journal={Journal of Cosmology and Astroparticle Physics},
   publisher={IOP Publishing},
   author={Chisari, Nora Elisa and Dvorkin, Cora},
   year={2013},
   month=dec, pages={029–029} }

@misc{vitório2025exploringcoregalaxyconnection,
      title={Exploring the Core-galaxy Connection}, 
      author={Isabele Souza Vitório and Michael Buehlmann and Eve Kovacs and Patricia Larsen and Nicholas Frontiere and Katrin Heitmann},
      year={2025},
      eprint={2407.00268},
      archivePrefix={arXiv},
      primaryClass={astro-ph.CO},
      url={https://arxiv.org/abs/2407.00268}, 
}

@misc{horowitz2022differentiablestochastichalooccupation,
      title={Differentiable Stochastic Halo Occupation Distribution}, 
      author={Benjamin Horowitz and ChangHoon Hahn and Francois Lanusse and Chirag Modi and Simone Ferraro},
      year={2022},
      eprint={2211.03852},
      archivePrefix={arXiv},
      primaryClass={astro-ph.CO},
      url={https://arxiv.org/abs/2211.03852}, 
}

@article{Lamman_2024,
   title={The IA Guide: A Breakdown of Intrinsic Alignment Formalisms},
   volume={7},
   ISSN={2565-6120},
   url={http://dx.doi.org/10.21105/astro.2309.08605},
   DOI={10.21105/astro.2309.08605},
   journal={The Open Journal of Astrophysics},
   publisher={Maynooth University},
   author={Lamman, Claire and Tsaprazi, Eleni and Shi, Jingjing and Šarčević, Nikolina Niko and Pyne, Susan and Legnani, Elisa and Ferreira, Tassia},
   year={2024},
   month=feb }

@article{krause16,
	Adsurl = {http://adsabs.harvard.edu/abs/2016MNRAS.456..207K},
	Archiveprefix = {arXiv},
	Author = {{Krause}, E. and {Eifler}, T. and {Blazek}, J.},
	Doi = {10.1093/mnras/stv2615},
	Eprint = {1506.08730},
	Journal = {\mnras},
	Month = feb,
	Pages = {207-222},
	Title = {{The impact of intrinsic alignment on current and future cosmic shear surveys}},
	Volume = 456,
	Year = 2016}

@misc{dvorkin2022machine,
      title={Machine Learning and Cosmology}, 
      author={Cora Dvorkin and Siddharth Mishra-Sharma and Brian Nord and V. Ashley Villar and Camille Avestruz and Keith Bechtol and Aleksandra Ćiprijanović and Andrew J. Connolly and Lehman H. Garrison and Gautham Narayan and Francisco Villaescusa-Navarro},
      year={2022},
      eprint={2203.08056},
      archivePrefix={arXiv},
      primaryClass={hep-ph}
}

@article{Zhai_2019,
   title={The Aemulus Project. III. Emulation of the Galaxy Correlation Function},
   volume={874},
   ISSN={1538-4357},
   url={http://dx.doi.org/10.3847/1538-4357/ab0d7b},
   DOI={10.3847/1538-4357/ab0d7b},
   number={1},
   journal={The Astrophysical Journal},
   publisher={American Astronomical Society},
   author={Zhai, Zhongxu and Tinker, Jeremy L. and Becker, Matthew R. and DeRose, Joseph and Mao, Yao-Yuan and McClintock, Thomas and McLaughlin, Sean and Rozo, Eduardo and Wechsler, Risa H.},
   year={2019},
   month=mar, pages={95} }

@article{Kwan_2023,
   title={Galaxy Clustering in the Mira-Titan Universe. I. Emulators for the Redshift Space Galaxy Correlation Function and Galaxy–Galaxy Lensing},
   volume={952},
   ISSN={1538-4357},
   url={http://dx.doi.org/10.3847/1538-4357/acd92f},
   DOI={10.3847/1538-4357/acd92f},
   number={1},
   journal={The Astrophysical Journal},
   publisher={American Astronomical Society},
   author={Kwan, Juliana and Saito, Shun and Leauthaud, Alexie and Heitmann, Katrin and Habib, Salman and Frontiere, Nicholas and Guo, Hong and Huang, Song and Pope, Adrian and Rodriguéz-Torres, Sergio},
   year={2023},
   month=jul, pages={80} }

@article{DUANE1987216,
title = {Hybrid Monte Carlo},
journal = {Physics Letters B},
volume = {195},
number = {2},
pages = {216-222},
year = {1987},
issn = {0370-2693},
doi = {https://doi.org/10.1016/0370-2693(87)91197-X},
url = {https://www.sciencedirect.com/science/article/pii/037026938791197X},
author = {Simon Duane and A.D. Kennedy and Brian J. Pendleton and Duncan Roweth}
}

@article{abacussummit,
   title={<scp>AbacusSummit</scp>: a massive set of high-accuracy, high-resolution N-body simulations},
   volume={508},
   ISSN={1365-2966},
   url={http://dx.doi.org/10.1093/mnras/stab2484},
   DOI={10.1093/mnras/stab2484},
   number={3},
   journal={Monthly Notices of the Royal Astronomical Society},
   publisher={Oxford University Press (OUP)},
   author={Maksimova, Nina A and Garrison, Lehman H and Eisenstein, Daniel J and Hadzhiyska, Boryana and Bose, Sownak and Satterthwaite, Thomas P},
   year={2021},
   month=sep, pages={4017–4037} }

@article{quijote,
   title={The Quijote Simulations},
   volume={250},
   ISSN={1538-4365},
   url={http://dx.doi.org/10.3847/1538-4365/ab9d82},
   DOI={10.3847/1538-4365/ab9d82},
   number={1},
   journal={The Astrophysical Journal Supplement Series},
   publisher={American Astronomical Society},
   author={Villaescusa-Navarro, Francisco and Hahn, ChangHoon and Massara, Elena and Banerjee, Arka and Delgado, Ana Maria and Ramanah, Doogesh Kodi and Charnock, Tom and Giusarma, Elena and Li, Yin and Allys, Erwan and Brochard, Antoine and Uhlemann, Cora and Chiang, Chi-Ting and He, Siyu and Pisani, Alice and Obuljen, Andrej and Feng, Yu and Castorina, Emanuele and Contardo, Gabriella and Kreisch, Christina D. and Nicola, Andrina and Alsing, Justin and Scoccimarro, Roman and Verde, Licia and Viel, Matteo and Ho, Shirley and Mallat, Stephane and Wandelt, Benjamin and Spergel, David N.},
   year={2020},
   month=aug, pages={2} }

@article{camels,
   title={The CAMELS Project: Cosmology and Astrophysics with Machine-learning Simulations},
   volume={915},
   ISSN={1538-4357},
   url={http://dx.doi.org/10.3847/1538-4357/abf7ba},
   DOI={10.3847/1538-4357/abf7ba},
   number={1},
   journal={The Astrophysical Journal},
   publisher={American Astronomical Society},
   author={Villaescusa-Navarro, Francisco and Anglés-Alcázar, Daniel and Genel, Shy and Spergel, David N. and S. Somerville, Rachel and Dave, Romeel and Pillepich, Annalisa and Hernquist, Lars and Nelson, Dylan and Torrey, Paul and Narayanan, Desika and Li, Yin and Philcox, Oliver and La Torre, Valentina and Maria Delgado, Ana and Ho, Shirley and Hassan, Sultan and Burkhart, Blakesley and Wadekar, Digvijay and Battaglia, Nicholas and Contardo, Gabriella and Bryan, Greg L.},
   year={2021},
   month=jul, pages={71} }

@article{Grand_n_2022,
   title={Differentiable Predictions for Large Scale Structure with SHAMNet},
   volume={5},
   ISSN={2565-6120},
   url={http://dx.doi.org/10.21105/astro.2205.11587},
   DOI={10.21105/astro.2205.11587},
   number={1},
   journal={The Open Journal of Astrophysics},
   publisher={Maynooth University},
   author={Grandón, Daniela and Sellentin, Elena},
   year={2022},
   month=jul }

@article{Paopiamsap_2024,
   title={Accuracy requirements on intrinsic alignments for Stage-IV cosmic shear},
   volume={7},
   ISSN={2565-6120},
   url={http://dx.doi.org/10.33232/001c.117419},
   DOI={10.33232/001c.117419},
   journal={The Open Journal of Astrophysics},
   publisher={Maynooth University},
   author={Paopiamsap, Anya and Porqueres, Natalia and Alonso, David and Harnois-Deraps, Joachim and Leonard, C. Danielle},
   year={2024},
   month=may }

@misc{pandey2024charmcreatinghalosautoregressive,
      title={CHARM: Creating Halos with Auto-Regressive Multi-stage networks}, 
      author={Shivam Pandey and Chirag Modi and Benjamin D. Wandelt and Deaglan J. Bartlett and Adrian E. Bayer and Greg L. Bryan and Matthew Ho and Guilhem Lavaux and T. Lucas Makinen and Francisco Villaescusa-Navarro},
      year={2024},
      eprint={2409.09124},
      archivePrefix={arXiv},
      primaryClass={astro-ph.CO},
      url={https://arxiv.org/abs/2409.09124}, 
}

@article{Jagvaral_2022,
   title={Galaxies and haloes on graph neural networks: Deep generative modelling scalar and vector quantities for intrinsic alignment},
   volume={516},
   ISSN={1365-2966},
   url={http://dx.doi.org/10.1093/mnras/stac2083},
   DOI={10.1093/mnras/stac2083},
   number={2},
   journal={Monthly Notices of the Royal Astronomical Society},
   publisher={Oxford University Press (OUP)},
   author={Jagvaral, Yesukhei and Lanusse, François and Singh, Sukhdeep and Mandelbaum, Rachel and Ravanbakhsh, Siamak and Campbell, Duncan},
   year={2022},
   month=aug, pages={2406–2419} }

@INPROCEEDINGS{MVE,
  author={Nix, D.A. and Weigend, A.S.},
  booktitle={Proceedings of 1994 IEEE International Conference on Neural Networks (ICNN'94)}, 
  title={Estimating the mean and variance of the target probability distribution}, 
  year={1994},
  volume={1},
  number={},
  pages={55-60 vol.1},
  doi={10.1109/ICNN.1994.374138}}

@misc{nelson2021illustristng,
      title={The IllustrisTNG Simulations: Public Data Release}, 
      author={Dylan Nelson and Volker Springel and Annalisa Pillepich and Vicente Rodriguez-Gomez and Paul Torrey and Shy Genel and Mark Vogelsberger and Ruediger Pakmor and Federico Marinacci and Rainer Weinberger and Luke Kelley and Mark Lovell and Benedikt Diemer and Lars Hernquist},
      year={2021},
      eprint={1812.05609},
      archivePrefix={arXiv},
      primaryClass={astro-ph.GA}
}

@article{Hearin_2017,
doi = {10.3847/1538-3881/aa859f},
url = {https://dx.doi.org/10.3847/1538-3881/aa859f},
year = {2017},
month = {oct},
publisher = {The American Astronomical Society},
volume = {154},
number = {5},
pages = {190},
author = {Andrew P. Hearin and Duncan Campbell and Erik Tollerud and Peter Behroozi and Benedikt Diemer and Nathan J. Goldbaum and Elise Jennings and Alexie Leauthaud and Yao-Yuan Mao and Surhud More and John Parejko and Manodeep Sinha and Brigitta Sipöcz and Andrew Zentner},
title = {Forward Modeling of Large-scale Structure: An Open-source Approach with Halotools},
journal = {The Astronomical Journal}}

@misc{xu2015empirical,
      title={Empirical Evaluation of Rectified Activations in Convolutional Network}, 
      author={Bing Xu and Naiyan Wang and Tianqi Chen and Mu Li},
      year={2015},
      eprint={1505.00853},
      archivePrefix={arXiv},
      primaryClass={cs.LG}
}

@misc{vanalfen_2023,
      title={An Empirical Model For Intrinsic Alignments: Insights From Cosmological Simulations}, 
      author={Nicholas {Van Alfen} and Duncan Campbell and Jonathan Blazek and C. Danielle Leonard and Francois Lanusse and Andrew Hearin and Rachel Mandelbaum and The LSST Dark Energy Science Collaboration},
      year={2024},
      eprint={2311.07374},
      archivePrefix={arXiv},
      primaryClass={astro-ph.CO}
}

@misc{loshchilov2019decoupled,
      title={Decoupled Weight Decay Regularization}, 
      author={Ilya Loshchilov and Frank Hutter},
      year={2019},
      eprint={1711.05101},
      archivePrefix={arXiv},
      primaryClass={cs.LG}
}

@misc{gal2016dropout,
      title={Dropout as a Bayesian Approximation: Representing Model Uncertainty in Deep Learning}, 
      author={Yarin Gal and Zoubin Ghahramani},
      year={2016},
      eprint={1506.02142},
      archivePrefix={arXiv},
      primaryClass={stat.ML}
}

@article{JMLR:v15:srivastava14a,
  author  = {Nitish Srivastava and Geoffrey Hinton and Alex Krizhevsky and Ilya Sutskever and Ruslan Salakhutdinov},
  title   = {Dropout: A Simple Way to Prevent Neural Networks from Overfitting},
  journal = {Journal of Machine Learning Research},
  year    = {2014},
  volume  = {15},
  number  = {56},
  pages   = {1929--1958},
  url     = {http://jmlr.org/papers/v15/srivastava14a.html}
}

@misc{kendall2018multitasklearningusinguncertainty,
      title={Multi-Task Learning Using Uncertainty to Weigh Losses for Scene Geometry and Semantics}, 
      author={Alex Kendall and Yarin Gal and Roberto Cipolla},
      year={2018},
      eprint={1705.07115},
      archivePrefix={arXiv},
      primaryClass={cs.CV},
      url={https://arxiv.org/abs/1705.07115}, 
}

@article{Zheng_2007,
	doi = {10.1086/521074},
	url = {https://doi.org/10.1086\%2F521074},
	year = 2007,
	month = {oct},
	publisher = {American Astronomical Society},
	volume = {667},
	number = {2},
	pages = {760--779},
	author = {Zheng Zheng and Alison L. Coil and Idit Zehavi},
	title = {Galaxy Evolution from Halo Occupation Distribution Modeling of {DEEP}2 and {SDSS} Galaxy Clustering},
	journal = {The Astrophysical Journal}
}

@article{TROXEL20151,
title = {The intrinsic alignment of galaxies and its impact on weak gravitational lensing in an era of precision cosmology},
journal = {Physics Reports},
volume = {558},
pages = {1-59},
year = {2015},
note = {The intrinsic alignment of galaxies and its impact on weak gravitational lensing in an era of precision cosmology},
issn = {0370-1573},
doi = {https://doi.org/10.1016/j.physrep.2014.11.001},
url = {https://www.sciencedirect.com/science/article/pii/S0370157314003974},
author = {M.A. Troxel and Mustapha Ishak}
}

@article{Bernstein_2002,
doi = {10.1086/338085},
url = {https://dx.doi.org/10.1086/338085},
year = {2002},
month = {feb},
publisher = {},
volume = {123},
number = {2},
pages = {583},
author = {G. M. Bernstein and M. Jarvis},
title = {Shapes and Shears, Stars and Smears: Optimal Measurements for Weak Lensing},
journal = {The Astronomical Journal}
}

@ARTICLE{2010ApJ...713.1322L,
       author = {{Lawrence}, Earl and {Heitmann}, Katrin and {White}, Martin and {Higdon}, David and {Wagner}, Christian and {Habib}, Salman and {Williams}, Brian},
        title = "{The Coyote Universe. III. Simulation Suite and Precision Emulator for the Nonlinear Matter Power Spectrum}",
      journal = {\apj},
         year = 2010,
        month = apr,
       volume = {713},
       number = {2},
        pages = {1322-1331},
          doi = {10.1088/0004-637X/713/2/1322},
archivePrefix = {arXiv},
       eprint = {0912.4490},
 primaryClass = {astro-ph.CO},
       adsurl = {https://ui.adsabs.harvard.edu/abs/2010ApJ...713.1322L}
}

@ARTICLE{2021arXiv210414568A,
       author = {{Aric{\`o}}, Giovanni and {Angulo}, Raul E. and {Zennaro}, Matteo},
        title = "{Accelerating Large-Scale-Structure data analyses by emulating Boltzmann solvers and Lagrangian Perturbation Theory}",
      journal = {arXiv e-prints},
         year = 2021,
        month = apr,
          eid = {arXiv:2104.14568},
        pages = {arXiv:2104.14568},
          doi = {10.48550/arXiv.2104.14568},
archivePrefix = {arXiv},
       eprint = {2104.14568},
 primaryClass = {astro-ph.CO},
       adsurl = {https://ui.adsabs.harvard.edu/abs/2021arXiv210414568A}
}

@ARTICLE{2021MNRAS.506.4070A,
       author = {{Aric{\`o}}, Giovanni and {Angulo}, Raul E. and {Contreras}, Sergio and {Ondaro-Mallea}, Lurdes and {Pellejero-Iba{\~n}ez}, Marcos and {Zennaro}, Matteo},
        title = "{The BACCO simulation project: a baryonification emulator with neural networks}",
      journal = {\mnras},
         year = 2021,
        month = sep,
       volume = {506},
       number = {3},
        pages = {4070-4082},
          doi = {10.1093/mnras/stab1911},
archivePrefix = {arXiv},
       eprint = {2011.15018},
 primaryClass = {astro-ph.CO},
       adsurl = {https://ui.adsabs.harvard.edu/abs/2021MNRAS.506.4070A}
}

@ARTICLE{2023arXiv231211707J,
       author = {{Jagvaral}, Yesukhei and {Lanusse}, Francois and {Mandelbaum}, Rachel},
        title = "{Unified framework for diffusion generative models in SO(3): applications in computer vision and astrophysics}",
      journal = {arXiv e-prints},
         year = 2023,
        month = dec,
          eid = {arXiv:2312.11707},
        pages = {arXiv:2312.11707},
          doi = {10.48550/arXiv.2312.11707},
archivePrefix = {arXiv},
       eprint = {2312.11707},
 primaryClass = {cs.LG},
       adsurl = {https://ui.adsabs.harvard.edu/abs/2023arXiv231211707J}
}

@misc{seitzer2022pitfallsheteroscedasticuncertaintyestimation,
      title={On the Pitfalls of Heteroscedastic Uncertainty Estimation with Probabilistic Neural Networks}, 
      author={Maximilian Seitzer and Arash Tavakoli and Dimitrije Antic and Georg Martius},
      year={2022},
      eprint={2203.09168},
      archivePrefix={arXiv},
      primaryClass={cs.LG},
      url={https://arxiv.org/abs/2203.09168}, 
}

@misc{sluijterman2023optimaltrainingmeanvariance,
      title={Optimal Training of Mean Variance Estimation Neural Networks}, 
      author={Laurens Sluijterman and Eric Cator and Tom Heskes},
      year={2023},
      eprint={2302.08875},
      archivePrefix={arXiv},
      primaryClass={stat.ML},
      url={https://arxiv.org/abs/2302.08875}, 
}

@misc{he2015deepresiduallearningimage,
      title={Deep Residual Learning for Image Recognition}, 
      author={Kaiming He and Xiangyu Zhang and Shaoqing Ren and Jian Sun},
      year={2015},
      eprint={1512.03385},
      archivePrefix={arXiv},
      primaryClass={cs.CV},
      url={https://arxiv.org/abs/1512.03385}, 
}

@ARTICLE{lsst,
       author = {{Ivezi{\'c}}, {\v{Z}}eljko and {Kahn}, Steven M. and {Tyson}, J. Anthony and {Abel}, Bob and {Acosta}, Emily and {Allsman}, Robyn and {Alonso}, David and {AlSayyad}, Yusra and {Anderson}, Scott F. and {Andrew}, John and {Angel}, James Roger P. and {Angeli}, George Z. and {Ansari}, Reza and {Antilogus}, Pierre and {Araujo}, Constanza and {Armstrong}, Robert and {Arndt}, Kirk T. and {Astier}, Pierre and {Aubourg}, {\'E}ric and {Auza}, Nicole and {Axelrod}, Tim S. and {Bard}, Deborah J. and {Barr}, Jeff D. and {Barrau}, Aurelian and {Bartlett}, James G. and {Bauer}, Amanda E. and {Bauman}, Brian J. and {Baumont}, Sylvain and {Bechtol}, Ellen and {Bechtol}, Keith and {Becker}, Andrew C. and {Becla}, Jacek and {Beldica}, Cristina and {Bellavia}, Steve and {Bianco}, Federica B. and {Biswas}, Rahul and {Blanc}, Guillaume and {Blazek}, Jonathan and {Blandford}, Roger D. and {Bloom}, Josh S. and {Bogart}, Joanne and {Bond}, Tim W. and {Booth}, Michael T. and {Borgland}, Anders W. and {Borne}, Kirk and {Bosch}, James F. and {Boutigny}, Dominique and {Brackett}, Craig A. and {Bradshaw}, Andrew and {Brandt}, William Nielsen and {Brown}, Michael E. and {Bullock}, James S. and {Burchat}, Patricia and {Burke}, David L. and {Cagnoli}, Gianpietro and {Calabrese}, Daniel and {Callahan}, Shawn and {Callen}, Alice L. and {Carlin}, Jeffrey L. and {Carlson}, Erin L. and {Chandrasekharan}, Srinivasan and {Charles-Emerson}, Glenaver and {Chesley}, Steve and {Cheu}, Elliott C. and {Chiang}, Hsin-Fang and {Chiang}, James and {Chirino}, Carol and {Chow}, Derek and {Ciardi}, David R. and {Claver}, Charles F. and {Cohen-Tanugi}, Johann and {Cockrum}, Joseph J. and {Coles}, Rebecca and {Connolly}, Andrew J. and {Cook}, Kem H. and {Cooray}, Asantha and {Covey}, Kevin R. and {Cribbs}, Chris and {Cui}, Wei and {Cutri}, Roc and {Daly}, Philip N. and {Daniel}, Scott F. and {Daruich}, Felipe and {Daubard}, Guillaume and {Daues}, Greg and {Dawson}, William and {Delgado}, Francisco and {Dellapenna}, Alfred and {de Peyster}, Robert and {de Val-Borro}, Miguel and {Digel}, Seth W. and {Doherty}, Peter and {Dubois}, Richard and {Dubois-Felsmann}, Gregory P. and {Durech}, Josef and {Economou}, Frossie and {Eifler}, Tim and {Eracleous}, Michael and {Emmons}, Benjamin L. and {Fausti Neto}, Angelo and {Ferguson}, Henry and {Figueroa}, Enrique and {Fisher-Levine}, Merlin and {Focke}, Warren and {Foss}, Michael D. and {Frank}, James and {Freemon}, Michael D. and {Gangler}, Emmanuel and {Gawiser}, Eric and {Geary}, John C. and {Gee}, Perry and {Geha}, Marla and {Gessner}, Charles J.~B. and {Gibson}, Robert R. and {Gilmore}, D. Kirk and {Glanzman}, Thomas and {Glick}, William and {Goldina}, Tatiana and {Goldstein}, Daniel A. and {Goodenow}, Iain and {Graham}, Melissa L. and {Gressler}, William J. and {Gris}, Philippe and {Guy}, Leanne P. and {Guyonnet}, Augustin and {Haller}, Gunther and {Harris}, Ron and {Hascall}, Patrick A. and {Haupt}, Justine and {Hernandez}, Fabio and {Herrmann}, Sven and {Hileman}, Edward and {Hoblitt}, Joshua and {Hodgson}, John A. and {Hogan}, Craig and {Howard}, James D. and {Huang}, Dajun and {Huffer}, Michael E. and {Ingraham}, Patrick and {Innes}, Walter R. and {Jacoby}, Suzanne H. and {Jain}, Bhuvnesh and {Jammes}, Fabrice and {Jee}, M. James and {Jenness}, Tim and {Jernigan}, Garrett and {Jevremovi{\'c}}, Darko and {Johns}, Kenneth and {Johnson}, Anthony S. and {Johnson}, Margaret W.~G. and {Jones}, R. Lynne and {Juramy-Gilles}, Claire and {Juri{\'c}}, Mario and {Kalirai}, Jason S. and {Kallivayalil}, Nitya J. and {Kalmbach}, Bryce and {Kantor}, Jeffrey P. and {Karst}, Pierre and {Kasliwal}, Mansi M. and {Kelly}, Heather and {Kessler}, Richard and {Kinnison}, Veronica and {Kirkby}, David and {Knox}, Lloyd and {Kotov}, Ivan V. and {Krabbendam}, Victor L. and {Krughoff}, K. Simon and {Kub{\'a}nek}, Petr and {Kuczewski}, John and {Kulkarni}, Shri and {Ku}, John and {Kurita}, Nadine R. and {Lage}, Craig S. and {Lambert}, Ron and {Lange}, Travis and {Langton}, J. Brian and {Le Guillou}, Laurent and {Levine}, Deborah and {Liang}, Ming and {Lim}, Kian-Tat and {Lintott}, Chris J. and {Long}, Kevin E. and {Lopez}, Margaux and {Lotz}, Paul J. and {Lupton}, Robert H. and {Lust}, Nate B. and {MacArthur}, Lauren A. and {Mahabal}, Ashish and {Mandelbaum}, Rachel and {Markiewicz}, Thomas W. and {Marsh}, Darren S. and {Marshall}, Philip J. and {Marshall}, Stuart and {May}, Morgan and {McKercher}, Robert and {McQueen}, Michelle and {Meyers}, Joshua and {Migliore}, Myriam and {Miller}, Michelle and {Mills}, David J. and {Miraval}, Connor and {Moeyens}, Joachim and {Moolekamp}, Fred E. and {Monet}, David G. and {Moniez}, Marc and {Monkewitz}, Serge and {Montgomery}, Christopher and {Morrison}, Christopher B. and {Mueller}, Fritz and {Muller}, Gary P. and {Mu{\~n}oz Arancibia}, Freddy and {Neill}, Douglas R. and {Newbry}, Scott P. and {Nief}, Jean-Yves and {Nomerotski}, Andrei and {Nordby}, Martin and {O'Connor}, Paul and {Oliver}, John and {Olivier}, Scot S. and {Olsen}, Knut and {O'Mullane}, William and {Ortiz}, Sandra and {Osier}, Shawn and {Owen}, Russell E. and {Pain}, Reynald and {Palecek}, Paul E. and {Parejko}, John K. and {Parsons}, James B. and {Pease}, Nathan M. and {Peterson}, J. Matt and {Peterson}, John R. and {Petravick}, Donald L. and {Libby Petrick}, M.~E. and {Petry}, Cathy E. and {Pierfederici}, Francesco and {Pietrowicz}, Stephen and {Pike}, Rob and {Pinto}, Philip A. and {Plante}, Raymond and {Plate}, Stephen and {Plutchak}, Joel P. and {Price}, Paul A. and {Prouza}, Michael and {Radeka}, Veljko and {Rajagopal}, Jayadev and {Rasmussen}, Andrew P. and {Regnault}, Nicolas and {Reil}, Kevin A. and {Reiss}, David J. and {Reuter}, Michael A. and {Ridgway}, Stephen T. and {Riot}, Vincent J. and {Ritz}, Steve and {Robinson}, Sean and {Roby}, William and {Roodman}, Aaron and {Rosing}, Wayne and {Roucelle}, Cecille and {Rumore}, Matthew R. and {Russo}, Stefano and {Saha}, Abhijit and {Sassolas}, Benoit and {Schalk}, Terry L. and {Schellart}, Pim and {Schindler}, Rafe H. and {Schmidt}, Samuel and {Schneider}, Donald P. and {Schneider}, Michael D. and {Schoening}, William and {Schumacher}, German and {Schwamb}, Megan E. and {Sebag}, Jacques and {Selvy}, Brian and {Sembroski}, Glenn H. and {Seppala}, Lynn G. and {Serio}, Andrew and {Serrano}, Eduardo and {Shaw}, Richard A. and {Shipsey}, Ian and {Sick}, Jonathan and {Silvestri}, Nicole and {Slater}, Colin T. and {Smith}, J. Allyn and {Smith}, R. Chris and {Sobhani}, Shahram and {Soldahl}, Christine and {Storrie-Lombardi}, Lisa and {Stover}, Edward and {Strauss}, Michael A. and {Street}, Rachel A. and {Stubbs}, Christopher W. and {Sullivan}, Ian S. and {Sweeney}, Donald and {Swinbank}, John D. and {Szalay}, Alexander and {Takacs}, Peter and {Tether}, Stephen A. and {Thaler}, Jon J. and {Thayer}, John Gregg and {Thomas}, Sandrine and {Thornton}, Adam J. and {Thukral}, Vaikunth and {Tice}, Jeffrey and {Trilling}, David E. and {Turri}, Max and {Van Berg}, Richard and {Vanden Berk}, Daniel and {Vetter}, Kurt and {Virieux}, Francoise and {Vucina}, Tomislav and {Wahl}, William and {Walkowicz}, Lucianne and {Walsh}, Brian and {Walter}, Christopher W. and {Wang}, Daniel L. and {Wang}, Shin-Yawn and {Warner}, Michael and {Wiecha}, Oliver and {Willman}, Beth and {Winters}, Scott E. and {Wittman}, David and {Wolff}, Sidney C. and {Wood-Vasey}, W. Michael and {Wu}, Xiuqin and {Xin}, Bo and {Yoachim}, Peter and {Zhan}, Hu},
        title = "{LSST: From Science Drivers to Reference Design and Anticipated Data Products}",
      journal = {\apj},
         year = 2019,
        month = mar,
       volume = {873},
       number = {2},
          eid = {111},
        pages = {111},
          doi = {10.3847/1538-4357/ab042c},
archivePrefix = {arXiv},
       eprint = {0805.2366},
 primaryClass = {astro-ph},
       adsurl = {https://ui.adsabs.harvard.edu/abs/2019ApJ...873..111I}
}

@article{euclid,
   title={Euclid preparation: I. The Euclid Wide Survey},
   volume={662},
   ISSN={1432-0746},
   url={http://dx.doi.org/10.1051/0004-6361/202141938},
   DOI={10.1051/0004-6361/202141938},
   journal={Astronomy &amp; Astrophysics},
   publisher={EDP Sciences},
   author={Scaramella, R. and Amiaux, J. and Mellier, Y. and Burigana, C. and Carvalho, C. S. and Cuillandre, J.-C. and Da Silva, A. and Derosa, A. and Dinis, J. and Maiorano, E. and Maris, M. and Tereno, I. and Laureijs, R. and Boenke, T. and Buenadicha, G. and Dupac, X. and Gaspar Venancio, L. M. and Gómez-Álvarez, P. and Hoar, J. and Lorenzo Alvarez, J. and Racca, G. D. and Saavedra-Criado, G. and Schwartz, J. and Vavrek, R. and Schirmer, M. and Aussel, H. and Azzollini, R. and Cardone, V. F. and Cropper, M. and Ealet, A. and Garilli, B. and Gillard, W. and Granett, B. R. and Guzzo, L. and Hoekstra, H. and Jahnke, K. and Kitching, T. and Maciaszek, T. and Meneghetti, M. and Miller, L. and Nakajima, R. and Niemi, S. M. and Pasian, F. and Percival, W. J. and Pottinger, S. and Sauvage, M. and Scodeggio, M. and Wachter, S. and Zacchei, A. and Aghanim, N. and Amara, A. and Auphan, T. and Auricchio, N. and Awan, S. and Balestra, A. and Bender, R. and Bodendorf, C. and Bonino, D. and Branchini, E. and Brau-Nogue, S. and Brescia, M. and Candini, G. P. and Capobianco, V. and Carbone, C. and Carlberg, R. G. and Carretero, J. and Casas, R. and Castander, F. J. and Castellano, M. and Cavuoti, S. and Cimatti, A. and Cledassou, R. and Congedo, G. and Conselice, C. J. and Conversi, L. and Copin, Y. and Corcione, L. and Costille, A. and Courbin, F. and Degaudenzi, H. and Douspis, M. and Dubath, F. and Duncan, C. A. J. and Dusini, S. and Farrens, S. and Ferriol, S. and Fosalba, P. and Fourmanoit, N. and Frailis, M. and Franceschi, E. and Franzetti, P. and Fumana, M. and Gillis, B. and Giocoli, C. and Grazian, A. and Grupp, F. and Haugan, S. V. H. and Holmes, W. and Hormuth, F. and Hudelot, P. and Kermiche, S. and Kiessling, A. and Kilbinger, M. and Kohley, R. and Kubik, B. and Kümmel, M. and Kunz, M. and Kurki-Suonio, H. and Lahav, O. and Ligori, S. and Lilje, P. B. and Lloro, I. and Mansutti, O. and Marggraf, O. and Markovic, K. and Marulli, F. and Massey, R. and Maurogordato, S. and Melchior, M. and Merlin, E. and Meylan, G. and Mohr, J. J. and Moresco, M. and Morin, B. and Moscardini, L. and Munari, E. and Nichol, R. C. and Padilla, C. and Paltani, S. and Peacock, J. and Pedersen, K. and Pettorino, V. and Pires, S. and Poncet, M. and Popa, L. and Pozzetti, L. and Raison, F. and Rebolo, R. and Rhodes, J. and Rix, H.-W. and Roncarelli, M. and Rossetti, E. and Saglia, R. and Schneider, P. and Schrabback, T. and Secroun, A. and Seidel, G. and Serrano, S. and Sirignano, C. and Sirri, G. and Skottfelt, J. and Stanco, L. and Starck, J. L. and Tallada-Crespí, P. and Tavagnacco, D. and Taylor, A. N. and Teplitz, H. I. and Toledo-Moreo, R. and Torradeflot, F. and Trifoglio, M. and Valentijn, E. A. and Valenziano, L. and Verdoes Kleijn, G. A. and Wang, Y. and Welikala, N. and Weller, J. and Wetzstein, M. and Zamorani, G. and Zoubian, J. and Andreon, S. and Baldi, M. and Bardelli, S. and Boucaud, A. and Camera, S. and Di Ferdinando, D. and Fabbian, G. and Farinelli, R. and Galeotta, S. and Graciá-Carpio, J. and Maino, D. and Medinaceli, E. and Mei, S. and Neissner, C. and Polenta, G. and Renzi, A. and Romelli, E. and Rosset, C. and Sureau, F. and Tenti, M. and Vassallo, T. and Zucca, E. and Baccigalupi, C. and Balaguera-Antolínez, A. and Battaglia, P. and Biviano, A. and Borgani, S. and Bozzo, E. and Cabanac, R. and Cappi, A. and Casas, S. and Castignani, G. and Colodro-Conde, C. and Coupon, J. and Courtois, H. M. and Cuby, J. and de la Torre, S. and Desai, S. and Dole, H. and Fabricius, M. and Farina, M. and Ferreira, P. G. and Finelli, F. and Flose-Reimberg, P. and Fotopoulou, S. and Ganga, K. and Gozaliasl, G. and Hook, I. M. and Keihanen, E. and Kirkpatrick, C. C. and Liebing, P. and Lindholm, V. and Mainetti, G. and Martinelli, M. and Martinet, N. and Maturi, M. and McCracken, H. J. and Metcalf, R. B. and Morgante, G. and Nightingale, J. and Nucita, A. and Patrizii, L. and Potter, D. and Riccio, G. and Sánchez, A. G. and Sapone, D. and Schewtschenko, J. A. and Schultheis, M. and Scottez, V. and Teyssier, R. and Tutusaus, I. and Valiviita, J. and Viel, M. and Vriend, W. and Whittaker, L.},
   year={2022},
   month=jun, pages={A112} }

@misc{berman2025softclustering,
      title={On Soft Clustering For Correlation Estimators: Model Uncertainty, Differentiability, and Surrogates}, 
      author={Edward Berman and Sneh Pandya and Jacqueline McCleary and Marko Shuntov and Caitlin Casey and Nicole Drakos and Andreas Faisst and Steven Gillman and Ghassem Gozaliasl and Natalie Hogg and Jeyhan Kartaltepe and Anton Koekemoer and Wilfried Mercier and Diana Scognamiglio and COSMOS-Web and : and The JWST Cosmic Origins Survey},
      year={2025},
      eprint={2504.06174},
      archivePrefix={arXiv},
      primaryClass={astro-ph.IM},
      url={https://arxiv.org/abs/2504.06174}, 
}

@misc{roman,
      title={The Wide Field Infrared Survey Telescope: 100 Hubbles for the 2020s}, 
      author={Rachel Akeson and Lee Armus and Etienne Bachelet and Vanessa Bailey and Lisa Bartusek and Andrea Bellini and Dominic Benford and David Bennett and Aparna Bhattacharya and Ralph Bohlin and Martha Boyer and Valerio Bozza and Geoffrey Bryden and Sebastiano Calchi Novati and Kenneth Carpenter and Stefano Casertano and Ami Choi and David Content and Pratika Dayal and Alan Dressler and Olivier Doré and S. Michael Fall and Xiaohui Fan and Xiao Fang and Alexei Filippenko and Steven Finkelstein and Ryan Foley and Steven Furlanetto and Jason Kalirai and B. Scott Gaudi and Karoline Gilbert and Julien Girard and Kevin Grady and Jenny Greene and Puragra Guhathakurta and Chen Heinrich and Shoubaneh Hemmati and David Hendel and Calen Henderson and Thomas Henning and Christopher Hirata and Shirley Ho and Eric Huff and Anne Hutter and Rolf Jansen and Saurabh Jha and Samson Johnson and David Jones and Jeremy Kasdin and Patrick Kelly and Robert Kirshner and Anton Koekemoer and Jeffrey Kruk and Nikole Lewis and Bruce Macintosh and Piero Madau and Sangeeta Malhotra and Kaisey Mandel and Elena Massara and Daniel Masters and Julie McEnery and Kristen McQuinn and Peter Melchior and Mark Melton and Bertrand Mennesson and Molly Peeples and Matthew Penny and Saul Perlmutter and Alice Pisani and Andrés Plazas and Radek Poleski and Marc Postman and Clément Ranc and Bernard Rauscher and Armin Rest and Aki Roberge and Brant Robertson and Steven Rodney and James Rhoads and Jason Rhodes and Russell Ryan Jr. au2 and Kailash Sahu and David Sand and Dan Scolnic and Anil Seth and Yossi Shvartzvald and Karelle Siellez and Arfon Smith and David Spergel and Keivan Stassun and Rachel Street and Louis-Gregory Strolger and Alexander Szalay and John Trauger and M. A. Troxel and Margaret Turnbull and Roeland van der Marel and Anja von der Linden and Yun Wang and David Weinberg and Benjamin Williams and Rogier Windhorst and Edward Wollack and Hao-Yi Wu and Jennifer Yee and Neil Zimmerman},
      year={2019},
      eprint={1902.05569},
      archivePrefix={arXiv},
      primaryClass={astro-ph.IM},
      url={https://arxiv.org/abs/1902.05569}, 
}

@ARTICLE{cooray_2002,
       author = {{Cooray}, Asantha and {Sheth}, Ravi},
        title = "{Halo models of large scale structure}",
      journal = {\physrep},
         year = 2002,
        month = dec,
       volume = {372},
       number = {1},
        pages = {1-129},
          doi = {10.1016/S0370-1573(02)00276-4},
archivePrefix = {arXiv},
       eprint = {astro-ph/0206508},
 primaryClass = {astro-ph},
       adsurl = {https://ui.adsabs.harvard.edu/abs/2002PhR...372....1C}
}

@ARTICLE{asgari_2023,
       author = {{Asgari}, Marika and {Mead}, Alexander J. and {Heymans}, Catherine},
        title = "{The halo model for cosmology: a pedagogical review}",
      journal = {The Open Journal of Astrophysics},
         year = 2023,
        month = nov,
       volume = {6},
          eid = {39},
        pages = {39},
          doi = {10.21105/astro.2303.08752},
archivePrefix = {arXiv},
       eprint = {2303.08752},
 primaryClass = {astro-ph.CO},
       adsurl = {https://ui.adsabs.harvard.edu/abs/2023OJAp....6E..39A}
}

@article{landy_1993,
       author = {{Landy}, Stephen D. and {Szalay}, Alexander S.},
        title = "{Bias and Variance of Angular Correlation Functions}",
      journal = {\apj},
         year = 1993,
        month = jul,
       volume = {412},
        pages = {64},
          doi = {10.1086/172900},
       adsurl = {https://ui.adsabs.harvard.edu/abs/1993ApJ...412...64L}
}

@article{sukhdeep_2017,
    author = {Singh, Sukhdeep and Mandelbaum, Rachel and Seljak, Uroš and Slosar, Anže and Vazquez Gonzalez, Jose},
    title = "{Galaxy–galaxy lensing estimators and their covariance properties}",
    journal = {Monthly Notices of the Royal Astronomical Society},
    volume = {471},
    number = {4},
    pages = {3827-3844},
    year = {2017},
    month = {07},
    issn = {0035-8711},
    doi = {10.1093/mnras/stx1828},
    url = {https://doi.org/10.1093/mnras/stx1828},
    eprint = {https://academic.oup.com/mnras/article-pdf/471/4/3827/19536722/stx1828.pdf},
}

@ARTICLE{Klypin_2011,
       author = {{Klypin}, Anatoly A. and {Trujillo-Gomez}, Sebastian and {Primack}, Joel},
        title = "{Dark Matter Halos in the Standard Cosmological Model: Results from the Bolshoi Simulation}",
      journal = {\apj},
         year = 2011,
        month = oct,
       volume = {740},
       number = {2},
          eid = {102},
        pages = {102},
          doi = {10.1088/0004-637X/740/2/102},
archivePrefix = {arXiv},
       eprint = {1002.3660},
 primaryClass = {astro-ph.CO},
       adsurl = {https://ui.adsabs.harvard.edu/abs/2011ApJ...740..102K}
}

@misc{hoffman2011nouturnsampleradaptivelysetting,
      title={The No-U-Turn Sampler: Adaptively Setting Path Lengths in Hamiltonian Monte Carlo}, 
      author={Matthew D. Hoffman and Andrew Gelman},
      year={2011},
      eprint={1111.4246},
      archivePrefix={arXiv},
      primaryClass={stat.CO},
      url={https://arxiv.org/abs/1111.4246}, 
}

@book{Goodfellow-et-al-2016,
    title={Deep Learning},
    author={Ian Goodfellow and Yoshua Bengio and Aaron Courville},
    publisher={MIT Press},
    note={\url{http://www.deeplearningbook.org}},
    year={2016}
}

@misc{
jagvaral2023diffusion,
title={{DIFFUSION} {GENERATIVE} {MODELS} {ON} {SO}(3)},
author={Yesukhei Jagvaral and Francois Lanusse and Rachel Mandelbaum},
year={2023},
url={https://openreview.net/forum?id=jHA-yCyBGb}
}

@misc{schanz2023stochasticsuperresolutioncosmologicalsimulations,
      title={Stochastic Super-resolution of Cosmological Simulations with Denoising Diffusion Models}, 
      author={Andreas Schanz and Florian List and Oliver Hahn},
      year={2023},
      eprint={2310.06929},
      archivePrefix={arXiv},
      primaryClass={astro-ph.CO},
      url={https://arxiv.org/abs/2310.06929}, 
}

@misc{mudur2024diffusionhmcparameterinferencediffusion,
      title={Diffusion-HMC: Parameter Inference with Diffusion Model driven Hamiltonian Monte Carlo}, 
      author={Nayantara Mudur and Carolina Cuesta-Lazaro and Douglas P. Finkbeiner},
      year={2024},
      eprint={2405.05255},
      archivePrefix={arXiv},
      primaryClass={astro-ph.CO},
      url={https://arxiv.org/abs/2405.05255}, 
}

@article{nbody,
author = {Perraudin, Nathanaël and Srivastava, Ankit and Lucchi, Aurelien and Kacprzak, Tomasz and Hofmann, Thomas and Réfrégier, Alexandre},
year = {2019},
month = {12},
pages = {},
title = {Cosmological N-body simulations: a challenge for scalable generative models},
volume = {6},
journal = {Computational Astrophysics and Cosmology},
doi = {10.1186/s40668-019-0032-1}
}

@article{Bridle_2007,
	Adsurl = {http://adsabs.harvard.edu/abs/2007NJPh....9..444B},
	Archiveprefix = {arXiv},
	Author = {{Bridle}, S. and {King}, L.},
	Doi = {10.1088/1367-2630/9/12/444},
	Eprint = {0705.0166},
	Journal = {New Journal of Physics},
	Month = dec,
	Pages = {444},
	Title = {{Dark energy constraints from cosmic shear power spectra: impact of intrinsic alignments on photometric redshift requirements}},
	Volume = 9,
	Year = 2007}

@article{Blazek_2019,
  title = {Beyond linear galaxy alignments},
  author = {Blazek, Jonathan A. and MacCrann, Niall and Troxel, M. A. and Fang, Xiao},
  journal = {Phys. Rev. D},
  volume = {100},
  issue = {10},
  pages = {103506},
  numpages = {19},
  year = {2019},
  month = {Nov},
  publisher = {American Physical Society},
  doi = {10.1103/PhysRevD.100.103506},
  url = {https://link.aps.org/doi/10.1103/PhysRevD.100.103506}
}

@ARTICLE{vlah_2020,
       author = {{Vlah}, Zvonimir and {Chisari}, Nora Elisa and {Schmidt}, Fabian},
        title = "{An EFT description of galaxy intrinsic alignments}",
      journal = {\jcap},
         year = 2020,
        month = jan,
       volume = {2020},
       number = {1},
          eid = {025},
        pages = {025},
          doi = {10.1088/1475-7516/2020/01/025},
archivePrefix = {arXiv},
       eprint = {1910.08085},
 primaryClass = {astro-ph.CO},
       adsurl = {https://ui.adsabs.harvard.edu/abs/2020JCAP...01..025V}
}

@ARTICLE{bakx_2023,
       author = {{Bakx}, Thomas and {Kurita}, Toshiki and {Elisa Chisari}, Nora and {Vlah}, Zvonimir and {Schmidt}, Fabian},
        title = "{Effective field theory of intrinsic alignments at one loop order: a comparison to dark matter simulations}",
      journal = {\jcap},
         year = 2023,
        month = oct,
       volume = {2023},
       number = {10},
          eid = {005},
        pages = {005},
          doi = {10.1088/1475-7516/2023/10/005},
archivePrefix = {arXiv},
       eprint = {2303.15565},
 primaryClass = {astro-ph.CO},
       adsurl = {https://ui.adsabs.harvard.edu/abs/2023JCAP...10..005B}
}

@ARTICLE{maion_2023,
       author = {{Maion}, Francisco and {Angulo}, Raul E. and {Bakx}, Thomas and {Chisari}, Nora Elisa and {Kurita}, Toshiki and {Pellejero-Ib{\'a}{\~n}ez}, Marcos},
        title = "{HYMALAIA: A Hybrid Lagrangian Model for Intrinsic Alignments}",
      journal = {arXiv e-prints},
         year = 2023,
        month = jul,
          eid = {arXiv:2307.13754},
        pages = {arXiv:2307.13754},
          doi = {10.48550/arXiv.2307.13754},
archivePrefix = {arXiv},
       eprint = {2307.13754},
 primaryClass = {astro-ph.CO},
       adsurl = {https://ui.adsabs.harvard.edu/abs/2023arXiv230713754M}
}

@ARTICLE{chen_2023,
       author = {{Chen}, Shi-Fan and {Kokron}, Nickolas},
        title = "{A Lagrangian theory for galaxy shape statistics}",
      journal = {arXiv e-prints},
         year = 2023,
        month = sep,
          eid = {arXiv:2309.16761},
        pages = {arXiv:2309.16761},
          doi = {10.48550/arXiv.2309.16761},
archivePrefix = {arXiv},
       eprint = {2309.16761},
 primaryClass = {astro-ph.CO},
       adsurl = {https://ui.adsabs.harvard.edu/abs/2023arXiv230916761C}
}

@article{Joachimi_2013,
    author = {Joachimi, B. and Semboloni, E. and Hilbert, S. and Bett, P. E. and Hartlap, J. and Hoekstra, H. and Schneider, P.},
    title = "{Intrinsic galaxy shapes and alignments – II. Modelling the intrinsic alignment contamination of weak lensing surveys}",
    journal = {Monthly Notices of the Royal Astronomical Society},
    volume = {436},
    number = {1},
    pages = {819-838},
    year = {2013},
    month = {09},
    issn = {0035-8711},
    doi = {10.1093/mnras/stt1618},
    url = {https://doi.org/10.1093/mnras/stt1618},
    eprint = {https://academic.oup.com/mnras/article-pdf/436/1/819/18501732/stt1618.pdf},
}

@ARTICLE{Hoffmann_2022,
       author = {{Hoffmann}, Kai and {Secco}, Lucas F. and {Blazek}, Jonathan and {Crocce}, Martin and {Tallada-Cresp{\'\i}}, Pau and {Samuroff}, Simon and {Prat}, Judit and {Carretero}, Jorge and {Fosalba}, Pablo and {Gazta{\~n}aga}, Enrique and {Castander}, Francisco J. and {DES Collaboration}},
        title = "{Modeling intrinsic galaxy alignment in the MICE simulation}",
      journal = {\prd},
         year = 2022,
        month = dec,
       volume = {106},
       number = {12},
          eid = {123510},
        pages = {123510},
          doi = {10.1103/PhysRevD.106.123510},
archivePrefix = {arXiv},
       eprint = {2206.14219},
 primaryClass = {astro-ph.CO},
       adsurl = {https://ui.adsabs.harvard.edu/abs/2022PhRvD.106l3510H}
}

@article{Blazek_2015,
doi = {10.1088/1475-7516/2015/08/015},
url = {https://dx.doi.org/10.1088/1475-7516/2015/08/015},
year = {2015},
month = {aug},
publisher = {},
volume = {2015},
number = {08},
pages = {015},
author = {Jonathan Blazek and Zvonimir Vlah and Uroš Seljak},
title = {Tidal alignment of galaxies},
journal = {Journal of Cosmology and Astroparticle Physics}
}

@ARTICLE{vlah_2021,
       author = {{Vlah}, Zvonimir and {Chisari}, Nora Elisa and {Schmidt}, Fabian},
        title = "{Galaxy shape statistics in the effective field theory}",
      journal = {\jcap},
         year = 2021,
        month = may,
       volume = {2021},
       number = {5},
          eid = {061},
        pages = {061},
          doi = {10.1088/1475-7516/2021/05/061},
archivePrefix = {arXiv},
       eprint = {2012.04114},
 primaryClass = {astro-ph.CO},
       adsurl = {https://ui.adsabs.harvard.edu/abs/2021JCAP...05..061V}
}

@misc{paszke2019pytorchimperativestylehighperformance,
      title={PyTorch: An Imperative Style, High-Performance Deep Learning Library}, 
      author={Adam Paszke and Sam Gross and Francisco Massa and Adam Lerer and James Bradbury and Gregory Chanan and Trevor Killeen and Zeming Lin and Natalia Gimelshein and Luca Antiga and Alban Desmaison and Andreas Köpf and Edward Yang and Zach DeVito and Martin Raison and Alykhan Tejani and Sasank Chilamkurthy and Benoit Steiner and Lu Fang and Junjie Bai and Soumith Chintala},
      year={2019},
      eprint={1912.01703},
      archivePrefix={arXiv},
      primaryClass={cs.LG},
      url={https://arxiv.org/abs/1912.01703}, 
}

@article{Nelson_2015,
   title={The illustris simulation: Public data release},
   volume={13},
   ISSN={2213-1337},
   url={http://dx.doi.org/10.1016/j.ascom.2015.09.003},
   DOI={10.1016/j.ascom.2015.09.003},
   journal={Astronomy and Computing},
   publisher={Elsevier BV},
   author={Nelson, D. and Pillepich, A. and Genel, S. and Vogelsberger, M. and Springel, V. and Torrey, P. and Rodriguez-Gomez, V. and Sijacki, D. and Snyder, G.F. and Griffen, B. and Marinacci, F. and Blecha, L. and Sales, L. and Xu, D. and Hernquist, L.},
   year={2015},
   month=nov, pages={12–37} }

@article{Pillepich_2017,
   title={First results from the IllustrisTNG simulations: the stellar mass content of groups and clusters of galaxies},
   volume={475},
   ISSN={1365-2966},
   url={http://dx.doi.org/10.1093/mnras/stx3112},
   DOI={10.1093/mnras/stx3112},
   number={1},
   journal={Monthly Notices of the Royal Astronomical Society},
   publisher={Oxford University Press (OUP)},
   author={Pillepich, Annalisa and Nelson, Dylan and Hernquist, Lars and Springel, Volker and Pakmor, Rüdiger and Torrey, Paul and Weinberger, Rainer and Genel, Shy and Naiman, Jill P and Marinacci, Federico and Vogelsberger, Mark},
   year={2017},
   month=dec, pages={648–675} }

@article{Springel_2017,
   title={First results from the IllustrisTNG simulations: matter and galaxy clustering},
   volume={475},
   ISSN={1365-2966},
   url={http://dx.doi.org/10.1093/mnras/stx3304},
   DOI={10.1093/mnras/stx3304},
   number={1},
   journal={Monthly Notices of the Royal Astronomical Society},
   publisher={Oxford University Press (OUP)},
   author={Springel, Volker and Pakmor, Rüdiger and Pillepich, Annalisa and Weinberger, Rainer and Nelson, Dylan and Hernquist, Lars and Vogelsberger, Mark and Genel, Shy and Torrey, Paul and Marinacci, Federico and Naiman, Jill},
   year={2017},
   month=dec, pages={676–698} }

@article{Nelson_2017,
   title={First results from the IllustrisTNG simulations: the galaxy colour bimodality},
   volume={475},
   ISSN={1365-2966},
   url={http://dx.doi.org/10.1093/mnras/stx3040},
   DOI={10.1093/mnras/stx3040},
   number={1},
   journal={Monthly Notices of the Royal Astronomical Society},
   publisher={Oxford University Press (OUP)},
   author={Nelson, Dylan and Pillepich, Annalisa and Springel, Volker and Weinberger, Rainer and Hernquist, Lars and Pakmor, Rüdiger and Genel, Shy and Torrey, Paul and Vogelsberger, Mark and Kauffmann, Guinevere and Marinacci, Federico and Naiman, Jill},
   year={2017},
   month=nov, pages={624–647} }

@article{Naiman_2018,
   title={First results from the IllustrisTNG simulations: a tale of two elements – chemical evolution of magnesium and europium},
   volume={477},
   ISSN={1365-2966},
   url={http://dx.doi.org/10.1093/mnras/sty618},
   DOI={10.1093/mnras/sty618},
   number={1},
   journal={Monthly Notices of the Royal Astronomical Society},
   publisher={Oxford University Press (OUP)},
   author={Naiman, Jill P and Pillepich, Annalisa and Springel, Volker and Ramirez-Ruiz, Enrico and Torrey, Paul and Vogelsberger, Mark and Pakmor, Rüdiger and Nelson, Dylan and Marinacci, Federico and Hernquist, Lars and Weinberger, Rainer and Genel, Shy},
   year={2018},
   month=mar, pages={1206–1224} }

@article{Marinacci_2018,
   title={First results from the IllustrisTNG simulations: radio haloes and magnetic fields},
   ISSN={1365-2966},
   url={http://dx.doi.org/10.1093/mnras/sty2206},
   DOI={10.1093/mnras/sty2206},
   journal={Monthly Notices of the Royal Astronomical Society},
   publisher={Oxford University Press (OUP)},
   author={Marinacci, Federico and Vogelsberger, Mark and Pakmor, Rüdiger and Torrey, Paul and Springel, Volker and Hernquist, Lars and Nelson, Dylan and Weinberger, Rainer and Pillepich, Annalisa and Naiman, Jill and Genel, Shy},
   year={2018},
   month=aug }

@article{Hullermeier_2021,
   title={Aleatoric and epistemic uncertainty in machine learning: an introduction to concepts and methods},
   volume={110},
   ISSN={1573-0565},
   url={http://dx.doi.org/10.1007/s10994-021-05946-3},
   DOI={10.1007/s10994-021-05946-3},
   number={3},
   journal={Machine Learning},
   publisher={Springer Science and Business Media LLC},
   author={Hüllermeier, Eyke and Waegeman, Willem},
   year={2021},
   month=mar, pages={457–506} }

@ARTICLE{pillepich18,
   author = {{Pillepich}, A. and {Springel}, V. and {Nelson}, D. and {Genel}, S. and 
	{Naiman}, J. and {Pakmor}, R. and {Hernquist}, L. and {Torrey}, P. and 
	{Vogelsberger}, M. and {Weinberger}, R. and {Marinacci}, F.},
    title = "{Simulating galaxy formation with the IllustrisTNG model}",
  journal = {\mnras},
archivePrefix = "arXiv",
   eprint = {1703.02970},
     year = 2018,
    month = jan,
   volume = 473,
    pages = {4077-4106},
      doi = {10.1093/mnras/stx2656},
   adsurl = {https://ui.adsabs.harvard.edu/abs/2018MNRAS.473.4077P}
}

@ARTICLE{delgado_2023,
       author = {{Delgado}, Ana Maria and {Hadzhiyska}, Boryana and {Bose}, Sownak and {Springel}, Volker and {Hernquist}, Lars and {Barrera}, Monica and {Pakmor}, R{\"u}diger and {Ferlito}, Fulvio and {Kannan}, Rahul and {Hern{\'a}ndez-Aguayo}, C{\'e}sar and {White}, Simon D.~M. and {Frenk}, Carlos},
        title = "{The MillenniumTNG project: intrinsic alignments of galaxies and haloes}",
      journal = {\mnras},
         year = 2023,
        month = aug,
       volume = {523},
       number = {4},
        pages = {5899-5914},
          doi = {10.1093/mnras/stad1781},
archivePrefix = {arXiv},
       eprint = {2304.12346},
 primaryClass = {astro-ph.CO},
       adsurl = {https://ui.adsabs.harvard.edu/abs/2023MNRAS.523.5899D}
}

@ARTICLE{jagvaral24_arxiv,
       author = {{Jagvaral}, Yesukhei and {Lanusse}, Francois and {Mandelbaum}, Rachel},
        title = "{Geometric deep learning for galaxy-halo connection: a case study for galaxy intrinsic alignments}",
      journal = {arXiv e-prints},
         year = 2024,
        month = sep,
          eid = {arXiv:2409.18761},
        pages = {arXiv:2409.18761},
          doi = {10.48550/arXiv.2409.18761},
archivePrefix = {arXiv},
       eprint = {2409.18761},
 primaryClass = {astro-ph.GA},
       adsurl = {https://ui.adsabs.harvard.edu/abs/2024arXiv240918761J}
}

@article{tenneti16,
	Adsurl = {http://adsabs.harvard.edu/abs/2016MNRAS.462.2668T},
	Archiveprefix = {arXiv},
	Author = {{Tenneti}, A. and {Mandelbaum}, R. and {Di Matteo}, T.},
	Doi = {10.1093/mnras/stw1823},
	Eprint = {1510.07024},
	Journal = {\mnras},
	Month = nov,
	Pages = {2668-2680},
	Title = {{Intrinsic alignments of disc and elliptical galaxies in the MassiveBlack-II and Illustris simulations}},
	Volume = 462,
	Year = 2016}

@ARTICLE{samuroff21,
       author = {{Samuroff}, S. and {Mandelbaum}, R. and {Blazek}, J.},
        title = "{Advances in constraining intrinsic alignment models with hydrodynamic simulations}",
      journal = {\mnras},
         year = 2021,
        month = nov,
       volume = {508},
       number = {1},
        pages = {637-664},
          doi = {10.1093/mnras/stab2520},
archivePrefix = {arXiv},
       eprint = {2009.10735},
       adsurl = {https://ui.adsabs.harvard.edu/abs/2021MNRAS.508..637S}
}

@ARTICLE{fortuna21a,
       author = {{Fortuna}, Maria Cristina and {Hoekstra}, Henk and {Joachimi}, Benjamin and {Johnston}, Harry and {Chisari}, Nora Elisa and {Georgiou}, Christos and {Mahony}, Constance},
        title = "{The halo model as a versatile tool to predict intrinsic alignments}",
      journal = {\mnras},
         year = 2021,
        month = feb,
       volume = {501},
       number = {2},
        pages = {2983-3002},
          doi = {10.1093/mnras/staa3802},
archivePrefix = {arXiv},
       eprint = {2003.02700},
 primaryClass = {astro-ph.CO},
       adsurl = {https://ui.adsabs.harvard.edu/abs/2021MNRAS.501.2983F}
}

@ARTICLE{secco22,
       author = {{Secco}, L.~F. and {Samuroff}, S. and {Krause}, E. and {Jain}, B. and {Blazek}, J. and {Raveri}, M. and {Campos}, A. and {Amon}, A. and {Chen}, A. and {Doux}, C. and {Choi}, A. and {Gruen}, D. and {Bernstein}, G.~M. and {Chang}, C. and {DeRose}, J. and {Myles}, J. and {Fert{\'e}}, A. and {Lemos}, P. and {Huterer}, D. and {Prat}, J. and {Troxel}, M.~A. and {MacCrann}, N. and {Liddle}, A.~R. and {Kacprzak}, T. and {Fang}, X. and {S{\'a}nchez}, C. and {Pandey}, S. and {Dodelson}, S. and {Chintalapati}, P. and {Hoffmann}, K. and {Alarcon}, A. and {Alves}, O. and {Andrade-Oliveira}, F. and {Baxter}, E.~J. and {Bechtol}, K. and {Becker}, M.~R. and {Brandao-Souza}, A. and {Camacho}, H. and {Carnero Rosell}, A. and {Carrasco Kind}, M. and {Cawthon}, R. and {Cordero}, J.~P. and {Crocce}, M. and {Davis}, C. and {Di Valentino}, E. and {Drlica-Wagner}, A. and {Eckert}, K. and {Eifler}, T.~F. and {Elidaiana}, M. and {Elsner}, F. and {Elvin-Poole}, J. and {Everett}, S. and {Fosalba}, P. and {Friedrich}, O. and {Gatti}, M. and {Giannini}, G. and {Gruendl}, R.~A. and {Harrison}, I. and {Hartley}, W.~G. and {Herner}, K. and {Huang}, H. and {Huff}, E.~M. and {Jarvis}, M. and {Jeffrey}, N. and {Kuropatkin}, N. and {Leget}, P. -F. and {Muir}, J. and {Mccullough}, J. and {Navarro Alsina}, A. and {Omori}, Y. and {Park}, Y. and {Porredon}, A. and {Rollins}, R. and {Roodman}, A. and {Rosenfeld}, R. and {Ross}, A.~J. and {Rykoff}, E.~S. and {Sanchez}, J. and {Sevilla-Noarbe}, I. and {Sheldon}, E.~S. and {Shin}, T. and {Troja}, A. and {Tutusaus}, I. and {Varga}, T.~N. and {Weaverdyck}, N. and {Wechsler}, R.~H. and {Yanny}, B. and {Yin}, B. and {Zhang}, Y. and {Zuntz}, J. and {Abbott}, T.~M.~C. and {Aguena}, M. and {Allam}, S. and {Annis}, J. and {Bacon}, D. and {Bertin}, E. and {Bhargava}, S. and {Bridle}, S.~L. and {Brooks}, D. and {Buckley-Geer}, E. and {Burke}, D.~L. and {Carretero}, J. and {Costanzi}, M. and {da Costa}, L.~N. and {De Vicente}, J. and {Diehl}, H.~T. and {Dietrich}, J.~P. and {Doel}, P. and {Ferrero}, I. and {Flaugher}, B. and {Frieman}, J. and {Garc{\'\i}a-Bellido}, J. and {Gaztanaga}, E. and {Gerdes}, D.~W. and {Giannantonio}, T. and {Gschwend}, J. and {Gutierrez}, G. and {Hinton}, S.~R. and {Hollowood}, D.~L. and {Honscheid}, K. and {Hoyle}, B. and {James}, D.~J. and {Jeltema}, T. and {Kuehn}, K. and {Lahav}, O. and {Lima}, M. and {Lin}, H. and {Maia}, M.~A.~G. and {Marshall}, J.~L. and {Martini}, P. and {Melchior}, P. and {Menanteau}, F. and {Miquel}, R. and {Mohr}, J.~J. and {Morgan}, R. and {Ogando}, R.~L.~C. and {Palmese}, A. and {Paz-Chinch{\'o}n}, F. and {Petravick}, D. and {Pieres}, A. and {Plazas Malag{\'o}n}, A.~A. and {Rodriguez-Monroy}, M. and {Romer}, A.~K. and {Sanchez}, E. and {Scarpine}, V. and {Schubnell}, M. and {Scolnic}, D. and {Serrano}, S. and {Smith}, M. and {Soares-Santos}, M. and {Suchyta}, E. and {Swanson}, M.~E.~C. and {Tarle}, G. and {Thomas}, D. and {To}, C. and {DES Collaboration}},
        title = "{Dark Energy Survey Year 3 results: Cosmology from cosmic shear and robustness to modeling uncertainty}",
      journal = {\prd},
         year = 2022,
        month = jan,
       volume = {105},
       number = {2},
          eid = {023515},
        pages = {023515},
          doi = {10.1103/PhysRevD.105.023515},
archivePrefix = {arXiv},
       eprint = {2105.13544},
 primaryClass = {astro-ph.CO},
       adsurl = {https://ui.adsabs.harvard.edu/abs/2022PhRvD.105b3515S}
}

@ARTICLE{campos23,
       author = {{Campos}, A. and {Samuroff}, S. and {Mandelbaum}, R.},
        title = "{An empirical approach to model selection: weak lensing and intrinsic alignments}",
      journal = {\mnras},
         year = 2023,
        month = oct,
       volume = {525},
       number = {2},
        pages = {1885-1901},
          doi = {10.1093/mnras/stad2213},
archivePrefix = {arXiv},
       eprint = {2211.02800},
 primaryClass = {astro-ph.CO},
       adsurl = {https://ui.adsabs.harvard.edu/abs/2023MNRAS.525.1885C}
}

@ARTICLE{samuroff24,
       author = {{Samuroff}, Simon and {Campos}, Andresa and {Porredon}, Anna and {Blazek}, Jonathan},
        title = "{Joint constraints from cosmic shear, galaxy-galaxy lensing and galaxy clustering: internal tension as an indicator of intrinsic alignment modelling error}",
      journal = {The Open Journal of Astrophysics},
         year = 2024,
        month = may,
       volume = {7},
          eid = {40},
        pages = {40},
          doi = {10.33232/001c.117964},
archivePrefix = {arXiv},
       eprint = {2402.15573},
 primaryClass = {astro-ph.CO},
       adsurl = {https://ui.adsabs.harvard.edu/abs/2024OJAp....7E..40S}
}

\newpage
\appendix

\section{HOD Model Construction}
\label{sec:appendix-hod-setup}

Constructing an HOD model with alignment requires us to choose an occupation model component to populate dark matter halos, a phase space model component to place these galaxies within their halo, and an alignment model component to assign an orientation to the galaxies.
Since we distinguish between central and satellite galaxies, we choose components for each of these steps for each of the two populations.
We choose the {\tt Zheng07Cens} and {\tt Zheng07Sats} occupation model components for central and satellite galaxies, respectively.
These components are available in {\tt halotools} and implement Equations 2, 3 and 5 from \citet{Zheng_2007}.
As discussed earlier, the five HOD parameters chosen for these models come from Table 1 of the same paper.

For simplicity, we use \texttt{TrivialPhaseSpace} and {\tt SubhaloPhaseSpace} as phase space model components for the central and satellite galaxies respectively, which places central galaxies at the location of their parent halo and satellite galaxies at the location of any subhalos (smaller dark matter halos that reside within the larger parent halo).
We use {\tt CentralAlignment} and {\tt RadialSatelliteAlignment} for the alignment model components.
The {\tt CentralAlignment} component aligns the central galaxy with respect to its parent halo, and {\tt RadialSatelliteAlignment} aligns the satellite galaxies with respect to the radial vector between the central galaxy and itself.
The parameters $\mu_{\rm cen}$ and $\mu_{\rm sat}$, the central and satellite alignment strengths, determine the shape of the Dimroth-Watson distribution from which the misalignment angles are drawn \citep{vanalfen_2023}.

\section{Correlation rescaling}
\label{sec:appendix-rescaling}

As \IAEmu outputs standardized correlations, it is crucial to properly rescale the model-predicted amplitudes, aleatoric uncertainties $\widehat{\sigma^{\text{aleo}}}$, and epistemic uncertainties $\widehat{\sigma^{\text{epi}}}$ for analysis.
For $\xi$, we denote $\hat{\xi}$ as the (standardized) model prediction, $\bar{\xi}$ refers to the $\xi$ correlations from the training dataset used for calculating statistics, and $\widehat{\sigma_\xi^{\text{aleo}}}$ and $\widehat{\sigma_\xi^{\text{epi}}}$ are the \IAEmu-predicted aleatoric and epistemic uncertainties for $\xi$.
The reverse transformation is as follows: 
\begin{align*}
    \xi &= \exp\left( \log \hat{\xi} \cdot \sigma_{\log \bar{\xi}} + \mu_{\log \bar{\xi}} \right).
\end{align*}
As a result of taking the $\log$ of $\xi$ for training, the rescaled aleatoric and epistemic uncertainties are only symmetric in log space.
The corresponding transformation of uncertainties from log to linear space follows from
the log normal moments:
\begin{align*}
    \sigma_{\xi}^{\text{aleo}} &= 
        \sqrt{\left( e^{(\sigma_{\log \xi}^{\text{aleo}})^2} - 1 \right)
        e^{2(\log \hat{\xi}\,\sigma_{\log \bar{\xi}} + \mu_{\log \bar{\xi}}) 
        + (\sigma_{\log \xi}^{\text{aleo}})^2}}, \\
    \sigma_{\xi}^{\text{epi}} &= 
        \exp\!\big(\log \hat{\xi}\,\sigma_{\log \bar{\xi}} + \mu_{\log \bar{\xi}}\big)
        \cdot \sigma_{\log \bar{\xi}} \cdot \widehat{\sigma_{\log \xi}^{\text{epi}}}.
\end{align*}
Here the aleatoric term follows from the exact log normal variance,
while the epistemic term is obtained by linear error propagation through the exponential mapping. 

\begin{figure*}
    \centering
    \begin{minipage}[t]{0.48\textwidth}
        \centering
        \includegraphics[width=\linewidth]{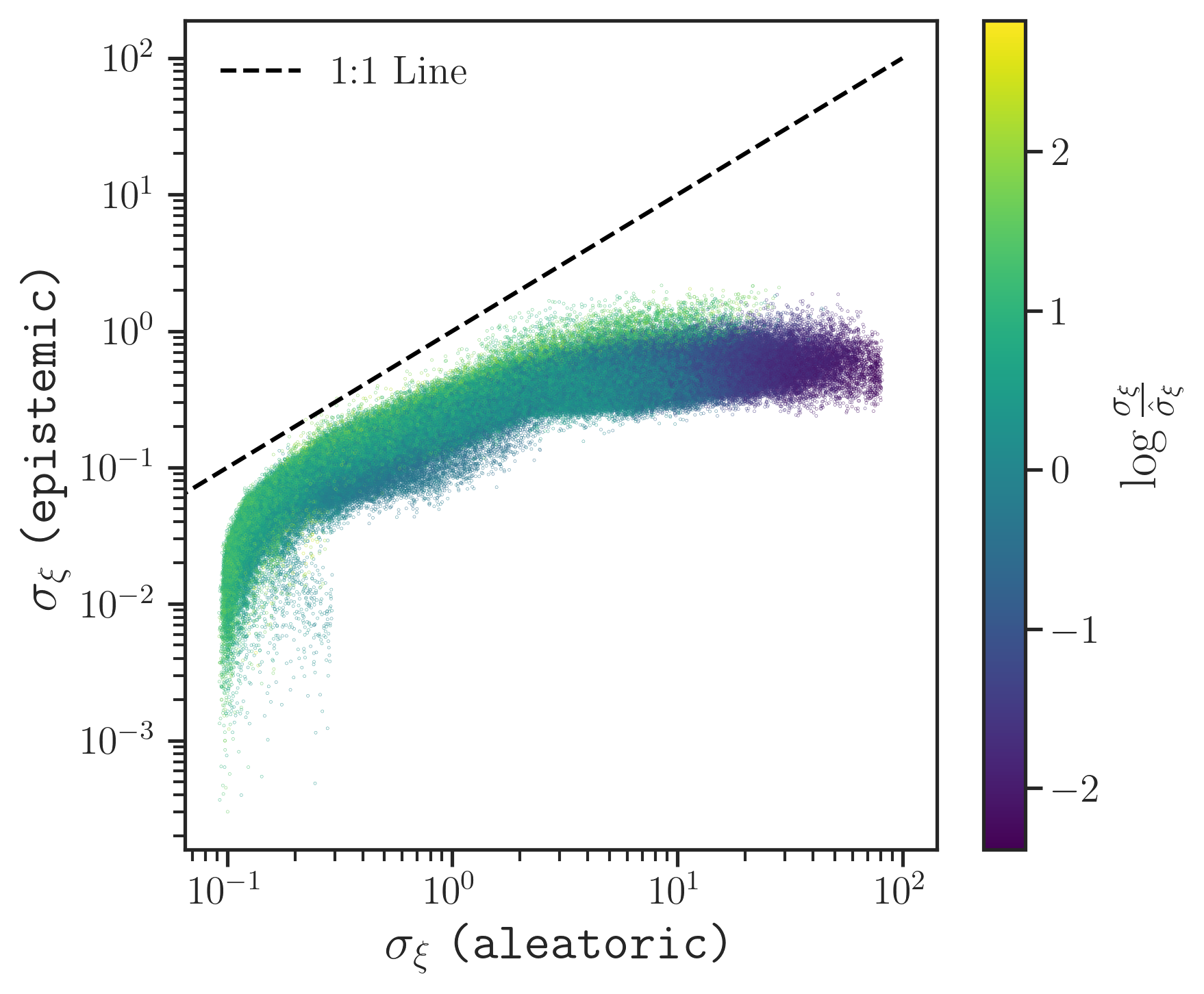}
    \end{minipage}\hfill
    \begin{minipage}[t]{0.48\textwidth}
        \centering
        \includegraphics[width=\linewidth]{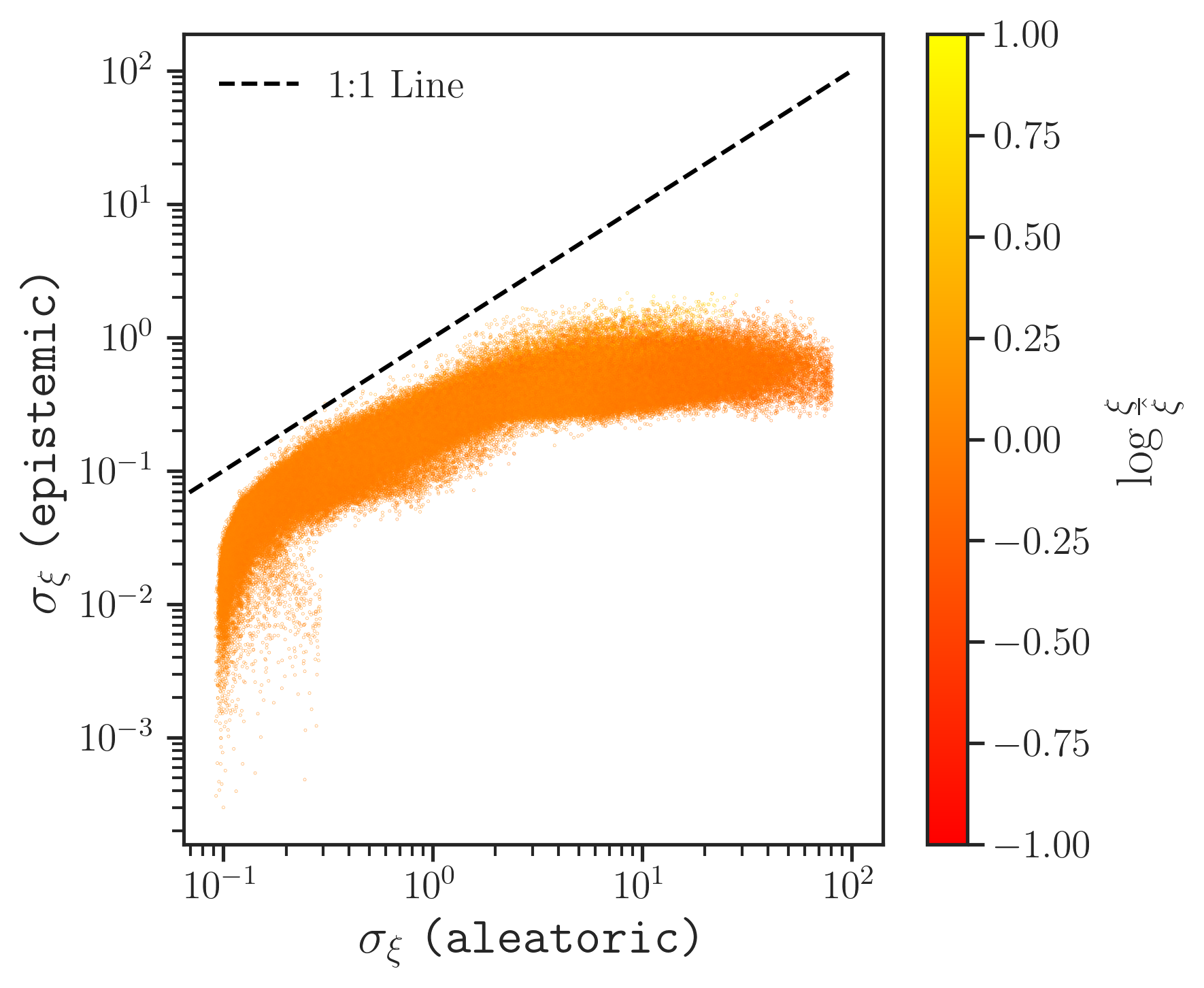}
    \end{minipage}
    \caption{Aleatoric vs. epistemic uncertainty comparison for $\xi$ with correlation uncertainty bias (left) and correlation amplitude bias (right). For test-set predictions, we analyze the total spread of aleatoric uncertainties of the data predicted by \IAEmu and epistemic uncertainties due to the stochasticity of \IAEmu. In the left plot, the coloring corresponds to the log-residual between \IAEmu predicted aleatoric uncertainties and (true) aleatoric uncertainties from \halotoolsia produced from the 10 realizations used in producing the dataset. In the right plot, the coloring corresponds to the log-residual between \IAEmu predicted correlation amplitudes and (mean) ground truth amplitudes from \halotoolsia produced from 10 realizations used in producing the dataset. It is seen that there is no clear correlations between log-residuals in the correlation amplitudes and \IAEmu aleatoric and epistemic uncertainties. 
    It is also seen that aleatoric uncertainties for $\xi$ are biased high when epistemic uncertainties are largest.}
    \label{fig:coverage_plots_xi}
\end{figure*}

We perform an analogous uncertainty analysis for $\xi$, following the methodology applied to $\omega$ and $\eta$ in Section \ref{sec:uncertainty}.
The results are presented in Figure \ref{fig:coverage_plots_xi}.
This analysis was omitted from the main discussion because $\xi$ exhibits a significantly higher signal correlation than $\omega$ and $\eta$, and the \IAEmu training procedure was not optimized to prioritize aleatoric uncertainty estimation for this quantity (see Section \ref{subsec:training}).
Nonetheless, the findings parallel those obtained for $\omega$ and $\eta$: the \IAEmu architecture appears sufficiently expressive for this task, as the scatter consistently lies below the 1:1 line.
We observe a median bias of $0.38$ dex in the aleatoric uncertainties, which exceeds the biases observed for $\omega$ and $\eta$.
This discrepancy likely stems from the choice of $\beta_\xi$, which was not tuned to facilitate accurate uncertainty quantification for $\xi$.
Consequently, \IAEmu tends to overestimate the aleatoric uncertainty when the epistemic uncertainty is large.
In contrast, a negligible median bias of $0.01$ dex is found for the correlation amplitudes.

The galaxy shape correlations $\omega$ and $\eta$ had no $\log$-scaling and therefore have a simpler inversion procedure:
\begin{align*}
\omega = \omega' \cdot \sigma_{\bar{\omega}} + \mu_{\bar{\omega}}, \quad \sigma_\omega^{\text{aleo}} = \sigma_{\bar{\omega}} \cdot \widehat{\sigma_{\omega}^{\text{aleo}}}, \quad \sigma_\omega^{\text{epi}} = \sigma_{\bar{\omega}} \cdot \widehat{\sigma_{\omega}^{\text{epi}}} \\
\eta = \eta' \cdot \sigma_{\bar{\eta}} + \mu_{\bar{\eta}}, \quad \sigma_\eta^{\text{aleo}} = \sigma_{\bar{\eta}} \cdot \widehat{\sigma_{\eta}^{\text{aleo}}}, \quad \sigma_\eta^{\text{epi}} = \sigma_{\bar{\eta}} \cdot \widehat{\sigma_{\eta}^{\text{epi}}}.
\end{align*}

\section{Training and Hamiltonian Monte Carlo}
\label{sec:appendix-training}

\textbf{Training.} We train the \IAEmu for a maximum of 500 epochs with a 100-epoch warm-up period and early stopping.
We also employ gradient clipping for numerical stability, as the training of MVE networks can suffer from instability.
The use of residual connections and a shallower embedding network than the decoder is to stabilize convergence during training.
We employ various techniques to further aid the convergence of the model.
Following the recommendations in \citep{sluijterman2023optimaltrainingmeanvariance}, we initialize all variance output-neurons to have a bias of zero which results in a constant variance prediction across all bins at initialization, ensuring that no bins are biased towards large variances. 

The choice of the $\beta$ parameter in the $\beta$-NLL loss (equation \ref{eq:NLL}) dictates the strength in which the loss interpolates between standard MSE and Gaussian-NLL loss.
The optimal value of $\beta$ will also not be the same for each correlation that is predicted.
We implement a warm-up period during training with $\beta = 1.0$ for all correlations to maximize regression on the means before transitioning to a value of $\beta_\xi = 0.9$ and $\beta_\omega = \beta_\eta = 0.5$ for the remainder of training.
The value of $\beta_\xi$ was chosen as $\xi(r)$ correlations exhibit a very high signal-to-noise ratio, so the aleatoric uncertainties on these correlations are generally not significant or of interest.  

We use the \texttt{AdamW} optimizer \citep{loshchilov2019decoupled} with a training batch size of 128 and a step learning rate scheduler (10\% decay at 167-epoch intervals with a starting learning rate of 0.01).
Additional L2-regularization via a weight decay factor of $10^{-4}$ is used in the optimizer.
All training was done on two NVIDIA A100-80GB GPUs.
During training, \IAEmu is validated every 5 epochs with an early stopping patience of 100 epochs based on the validation criteria.
The validation criteria for saving the model is a linear combination of MSE and Gaussian-NLL losses computed for each correlation \( \xi \), \( \omega \), and \( \eta \). 

The total MSE and NLL losses are calculated as the sum of each correlation, and the averaged validation losses are computed over the validation dataset.
The final combined validation loss \( \mathcal{L}_{\text{val}} \) is defined as:
\[
\mathcal{L}_{\text{val}} = \alpha \cdot \mathcal{L}_{\text{MSE}} + (1 - \alpha) \cdot \mathcal{L}_{\text{NLL}}
\]
where \( \alpha = 0.7 \) determines the weighting between MSE and NLL, guiding model selection based on this combined criterion.

\textbf{Hamiltonian Monte Carlo.} HMC is a variant of the Metropolis–Hastings algorithm, where Hamiltonian dynamics are simulated using a time-reversible, volume-preserving numerical integrator to propose transitions to new points in the state space.
We use HMC to sample from a posterior distribution over the inputs $x$, given trained NN parameters $\theta$ and observations $\mathcal{D}$.
This is described by
\begin{equation}
\label{eqn:bayes}
    p(x | \mathcal{D}, \theta) \propto p(\mathcal{D} | x, \theta) p(x) \;,
\end{equation}
Equation \ref{eqn:bayes} is a form of Bayes' Theorem, where $p(\mathcal{D} | x, \theta)$ is the likelihood function and $p(x)$ is the prior distribution on $x$.
HMC achieves this by forward modeling the dynamics of a governing Hamiltonian $H$:
\begin{equation}
    H = T + U = \frac{1}{2} \mathbf{p}^T M^{-1} \mathbf{p} - \ln p(x | \mathcal{D}, \theta)
\end{equation}
where $T$ is the kinetic energy with mass matrix $M$ and momentum $\mathbf{p}$, which controls the exploration in parameter space, and $-\ln p(x | \mathcal{D}, \theta)$ takes the role of the potential energy $U$.
The time-evolution of $x$  and $p$ is accordingly governed by Hamilton's equations:
\begin{equation}
\label{eqn:hamilton}
    \frac{d x}{d t} = \frac{\partial H}{\partial \mathbf{p}}, 
    \qquad
    \frac{d \mathbf{p}}{d t} = -\frac{\partial H}{\partial x}.
\end{equation}
HMC thus arrives at the posterior distribution over the inputs by sequentially evolving the dynamical variables according to Hamiltonian dynamics; this of course corresponds to minimizing the potential energy, which maximizes the log probability.
As seen in equation \ref{eqn:hamilton}, Hamilton's equations require gradients with respect to $H$, specifically $-\nabla_{x} \ln p(x | \mathcal{D}, \theta)$.
Decomposing this with chain rule, 
\begin{align}
\nabla_x \ln p(x | \mathcal{D}, \theta) &= \nabla_x \ln p(\mathcal{D} | x, \theta) + \nabla_x \ln p(x) \nonumber \\
&\propto \nabla_x \ln p(f_\theta(x) | \mathcal{D}) \nonumber \\
                                        &= \nabla_{f_\theta(x)} \ln p(f_\theta(x) | \mathcal{D}) \cdot \nabla_x f_\theta(x) \;, \nonumber
\end{align}
where in the second line we recognize that the likelihood is implicitly a function of the outputs of \IAEmu, $f_{\theta}(x)$, explicitly denoting its dependence on parameters $\theta$.
We thus see how  differentiability through the forward model is leveraged in this algorithm. 

\section{Example \IAEmu Predictions}
\label{sec:appendix-plots}

\begin{figure*}
    \centering
    \includegraphics[width=\textwidth]{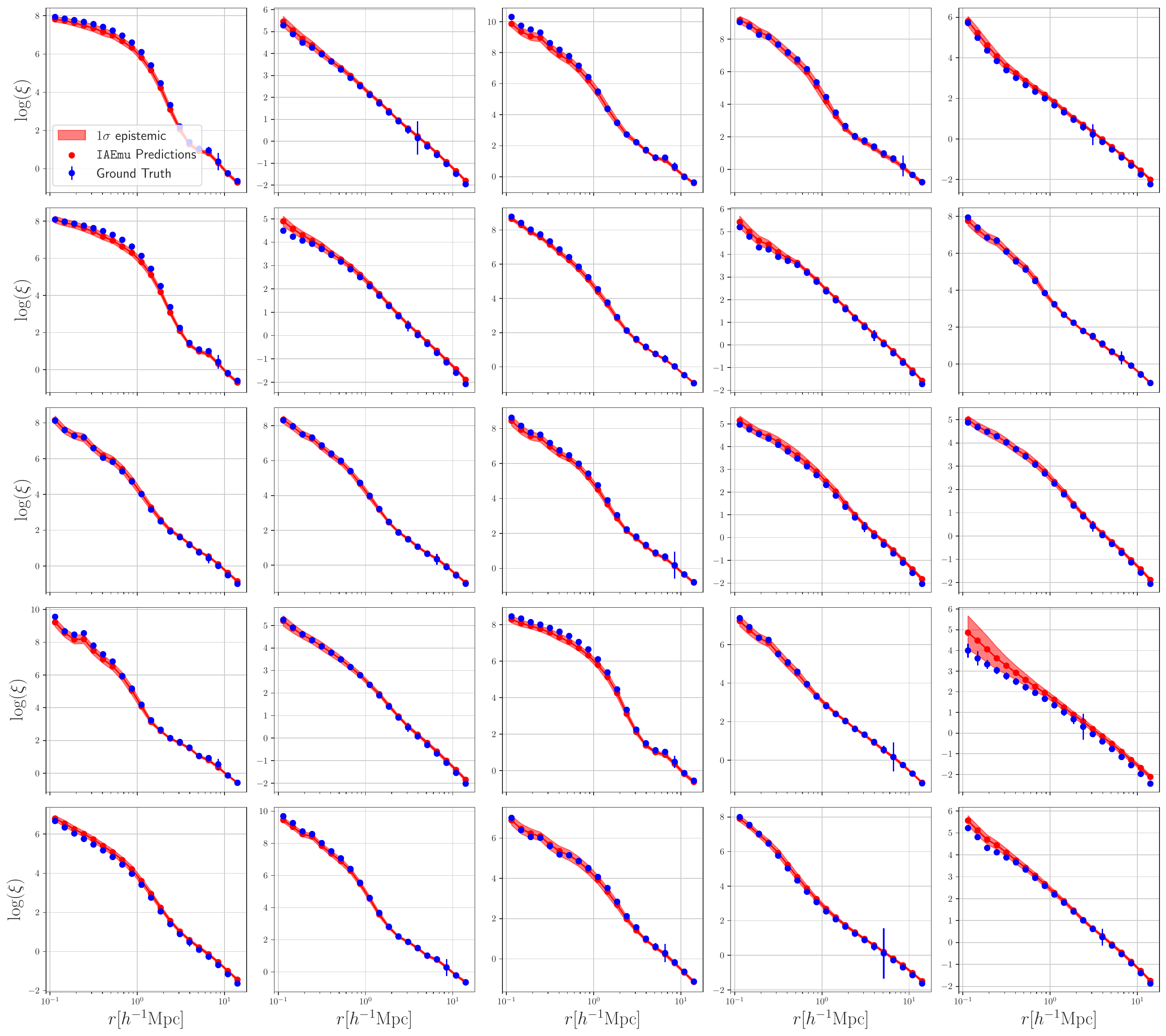}
    \caption{25 random \IAEmu test set predictions (red) compared with \halotoolsia ground truth (blue) for $\xi$. $\log (\xi)$ is plotted due to the large range in correlation amplitudes. Error bars on ground truth values are computed across 10 realizations of \protect\halotoolsia. $1 \sigma$ epistemic uncertainty for \IAEmu is shown in the red shaded region. The general agreement between \IAEmu and \halotoolsia for $\xi$ is observed.}
    \label{fig:xiappendix}
\end{figure*}

\begin{figure*}
    \centering
    \includegraphics[width=\textwidth]{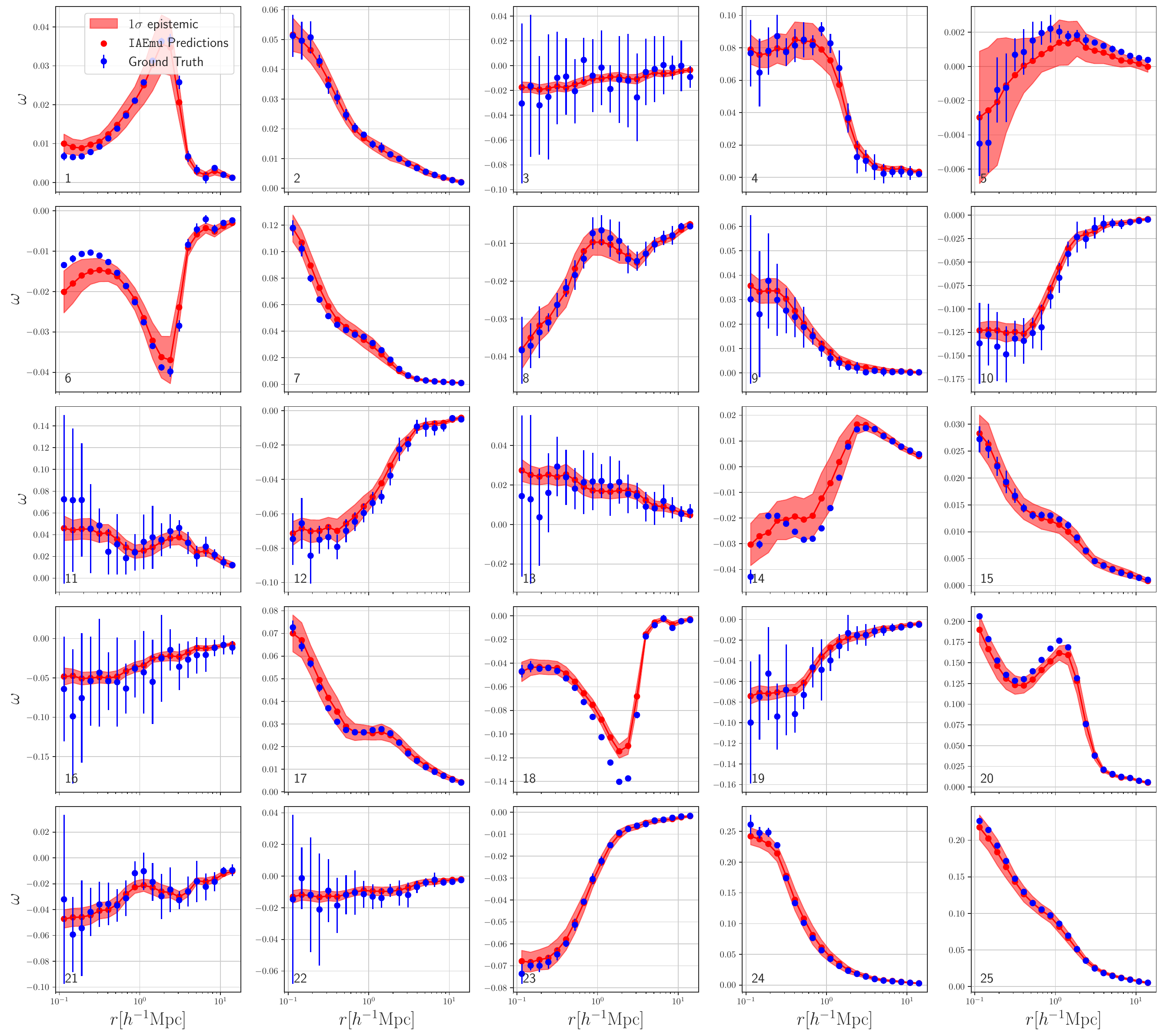}
    \caption{25 random \IAEmu test set predictions (red) compared with \halotoolsia ground truth (blue) for $\omega$. Error bars on ground truth values are computed across 10 realizations of \halotoolsia. $1 \sigma$ epistemic uncertainty for \IAEmu is shown in the red shaded region. The general agreement between \IAEmu and \halotoolsia for $\omega$ is observed, as well as the stochasticity, zero-crossings, and small amplitudes of the correlation. As mentioned in Section \ref{sec:performance}, these features make quantifying \IAEmu performance in terms of fractional error difficult.}
    \label{fig:omegaappendix}
\end{figure*}

\begin{figure*}
    \centering
    \includegraphics[width=\textwidth]{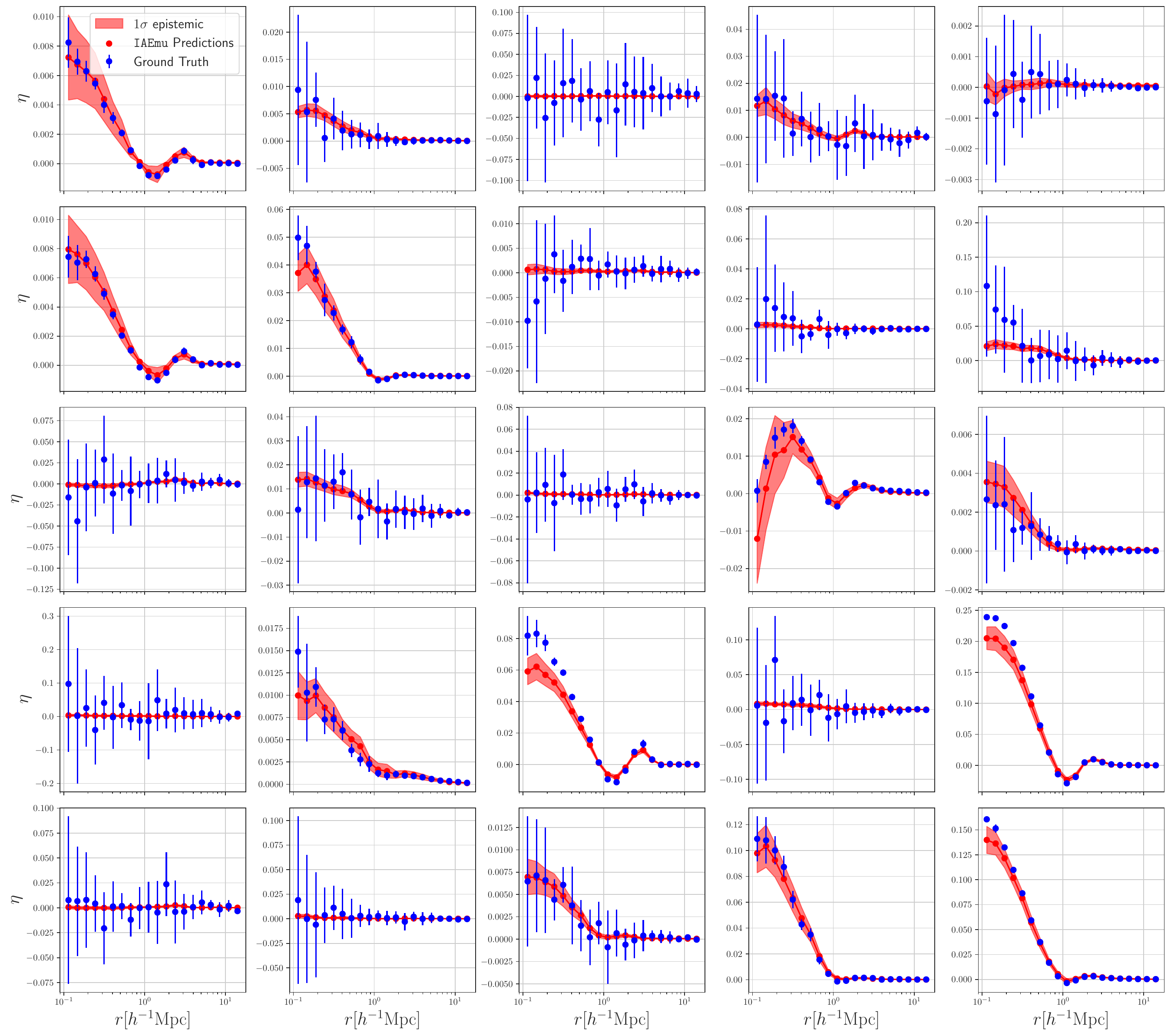}
    \caption{25 random \IAEmu test set predictions (red) compared with \halotoolsia ground truth (blue) for $\eta$. Error bars on ground truth values are computed across 10 realizations of \halotoolsia. $1 \sigma$ epistemic uncertainty for \IAEmu is shown in the red shaded region. The general agreement between \IAEmu and \halotoolsia for $\eta$ is observed, as well as the stochasticity, zero-crossings, and small amplitudes of the correlation. As mentioned in Section \ref{sec:performance}, these features make quantifying \IAEmu performance in terms of fractional error difficult.}
    \label{fig:etaappendix}
\end{figure*}

Randomly chosen \IAEmu test-set predictions comparing \halotoolsia and \IAEmu are shown in Figures \ref{fig:xiappendix}, \ref{fig:omegaappendix}, and \ref{fig:etaappendix}.
In the figures, the \halotoolsia ground truth is shown in blue, and \IAEmu with $1\sigma$ epistemic uncertainties are given in red.

\vspace{2 mm}
\end{document}